\documentclass[prl,reprint,showkeys,twocolumn,showpacs,preprintnumbers,amsmath,amssymb,nofootinbib,superscriptaddress]{revtex4-1}
\usepackage{graphicx}

\usepackage{breqn}

\usepackage{epsfig}
\usepackage{color}
\usepackage{amssymb}
\usepackage{amsmath}
\usepackage{mathrsfs}
\usepackage{hyperref}
\usepackage{multirow}

\makeatletter
\let\cat@comma@active\@empty
\makeatother

\begin{document}

\title{Hidden Analytic Relations for Two-Loop Higgs Amplitudes in QCD}
\author{Qingjun Jin}
\email{qjin@gscaep.ac.cn}
\affiliation{CAS Key Laboratory of Theoretical Physics, Institute of Theoretical Physics, Chinese Academy of Sciences, Beijing 100190, China}
\affiliation{Graduate School of China Academy of Engineering Physics, No. 10 Xibeiwang East Road, Haidian District, Beijing, 100193, China}

\author{Gang Yang}
\email{yangg@itp.ac.cn}
\affiliation{CAS Key Laboratory of Theoretical Physics, Institute of Theoretical Physics, Chinese Academy of Sciences, Beijing 100190, China}
\affiliation{School of Physical Sciences, University of Chinese Academy of Sciences,  No. 19A Yuquan Road, Beijing 100049, China}

\begin{abstract}

We compute the Higgs plus two-quark and one-gluon amplitudes ($H \rightarrow q \bar{q} g$) and  Higgs plus three-gluon amplitudes  ($H \rightarrow 3g$) in the Higgs effective theory with a general class of operators. By changing the quadratic Casimir $C_F$ to $C_A$, the maximally transcendental parts of the $H \rightarrow q \bar{q} g$ amplitudes turn out to be equivalent to that of the $H \rightarrow 3g$ amplitudes, which also coincide with the counterparts in ${\cal N}=4$ SYM. This generalizes the so-called maximal transcendentality principle to the Higgs amplitudes with external quark states, thus to the full QCD theory. We further verify that the correspondence applies also to two-loop form factors of more general operators, in both QCD and scalar-YM theory. Another interesting relation is also observed between the planar $H \rightarrow q \bar{q} g$ amplitudes and the minimal density form factors in ${\cal N}=4$ SYM.

\end{abstract}


\maketitle

\section{Introductioin}

Analytic studies have been crucial for uncovering new hidden structures of scattering amplitudes.
A famous example is the Parke-Taylor formula for tree-level maximally helicity violating (MHV) gluon amplitudes \cite{Parke:1986gb}, which is remarkably simple and hard to understand from the traditional Feynman diagram viewpoint.  
Another striking example is the six-point two-loop MHV amplitude in the planar ${\cal N}=4$ supersymmetric Yang-Mills theory (SYM), for which the original seventeen-page result \cite{DelDuca:2010zg} in terms of multiple polylogarithms turned out to be equivalent to a few lines of classical polylogarithms \cite{Goncharov:2010jf}.

While significant progress has been made in the study of loop amplitudes in supersymmetric theories such as ${\cal N}=4$ SYM, it should be fair to say that the analytic structures of amplitudes in realistic theories such as QCD are still largely unexplored beyond one loop. This is partially because the two-loop computation is itself very challenging in QCD and so far not many analytic results have been obtained. (See e.g.~\cite{Gehrmann:2015bfy, Dunbar:2016gjb, Dunbar:2017nfy, Badger:2018enw, Abreu:2018zmy, Abreu:2019odu} for recent progress on gluon amplitudes.) 
In this paper, we will study the two-loop Higgs amplitudes in the Higgs effective theory \cite{Wilczek:1977zn, Shifman:1979eb, Dawson:1990zj, Djouadi:1991tka, Kniehl:1995tn, Chetyrkin:1997sg, Chetyrkin:1997un} with full QCD corrections. These observables not only have a wide range of phenomenological applications, such as for the Higgs production at the Large Hadron Collider, see e.g.~\cite{Boughezal:2013uia,Chen:2014gva,Boughezal:2015dra,Boughezal:2015aha, Anastasiou:2016cez, Harlander:2016hcx, Anastasiou:2016hlm, Chen:2016zka, Lindert:2018iug,Jones:2018hbb,Neumann:2018bsx}, but are also important from a purely theoretical point of view, in particular, for uncovering hidden analytic structures of QCD amplitudes.
As a central result of this paper, we will show that the  so-called maximal transcendentality principle applies to both the Higgs plus 2-quark and 1-gluon amplitudes ($H \rightarrow q \bar q g$) and Higgs plus three-gluon amplitudes ($H \rightarrow 3g$).

The maximal transcendentality principle was conjectured first in \cite{Kotikov:2002ab, Kotikov:2004er} and states that, for certain quantities such as the anomalous dimensions, the QCD results \cite{Moch:2004pa} with highest transcendentality degree coincide with the ${\cal N}=4$ SYM results. 
Here, {\it transcendentality degree} characterizes the algebraic complexity of transcendental numbers or functions. For example, $\pi$ and $\log(x)$ have degree $1$, the Riemann zeta value $\zeta_n$ and polylogrithm function ${\rm Li}_n$ have degree $n$, and the transcendentality degree of algebraic numbers or rational functions is zero.
In \cite{Brandhuber:2012vm}, a further surprising observation was made: the two-loop BPS form factor in ${\cal N}=4$ SYM coincides with the maximally transcendental part of the $H \rightarrow 3g$ amplitudes with the dimension-5 operator $H{\rm tr}(F^2)$ \cite{Gehrmann:2011aa}. 
This implies that the maximal transcendentality principle does not only apply to pure numbers (such as anomalous dimensions) but also to kinematics-dependent functions (such as amplitudes).
More recently, the same correspondence has also found for the $H \rightarrow 3g$ amplitudes with higher dimension-7 operators and the corresponding form factors in ${\cal N}=4$ SYM \cite{Jin:2018fak, Brandhuber:2017bkg, Brandhuber:2018xzk, Brandhuber:2018kqb}. 
See other examples for Wilson lines \cite{Li:2014afw, Li:2016ctv} and an application for the collinear anomalous dimension \cite{Dixon:2017nat}.

So far the correspondence for Higgs amplitudes and form factors has been known for the cases with pure external gluon states. On the other hand, in full QCD, there are fundamental particles (i.e.~quarks). 
It is therefore very interesting to ask whether the maximal transcendentality principle applies to the Higgs amplitudes with external quarks as well.
This is a priori not obvious at all, since the quarks and gluons have very different color structures.
The surprising new observation of this paper is that the correspondence can be indeed extended to the $H \rightarrow q \bar q g$ amplitudes. 
Concretely, by converting the representation of quarks from fundamental to adjoint, the maximally transcendental parts of  the $H \rightarrow q \bar q g$ amplitudes become precisely the same as the corresponding $H \rightarrow 3g$ results. Such a generalized maximal transcendentality principle can be summarized as:
\begin{align}
& \  \textrm{Max. Tran. of } {(H \rightarrow q \bar q g)}|_{C_F \rightarrow C_A}  \nonumber \\
= & \  \textrm{Max. Tran. of } {(H \rightarrow 3g)} \nonumber \\
= & \  \textrm{Max. Tran. of ${\cal N}=4$ form factors} \nonumber \,.
\end{align}

We have checked this correspondence by computing several new results, including Higgs amplitudes with higher dimension operators in effective theory, as well as form factors with more general operators such as $\bar\psi\psi$. We also compute form factors involving fundamental scalar particles in the scalar-YM theory. They all satisfy the above correspondence, suggesting the relation is quite universal. 

As another interesting observation, we find that for the large $N_c$ limit of $H \rightarrow q \bar q g$ amplitudes with length-3 operators, the maximally transcendental part coincides with the minimal form factor density of higher length operators in ${\cal N}=4$ SYM \cite{Brandhuber:2014ica, Loebbert:2015ova, Brandhuber:2016fni, Loebbert:2016xkw}, up to simple $\zeta_3, \zeta_4$ terms. (The length of operator will be defined in the next section.) 
This suggests that the maximal transcendentality principle applies also to general minimal form factors with higher length operators in QCD.

In the next section, we will discuss the operators in the Higgs effective theory and briefly explain the computation of their form factors. In the further three sections, we explain in detail the maximal transcendentality principle, for both length-2 and length-3 cases, respectively. Finally, we give a summary and discussion.
The main results of this paper are summarized in Table~\ref{tab:summary}.

\section{Setup and Computation}

The dominant Higgs production at the LHC is the gluon fusion through a top quark loop. The corresponding Higgs amplitudes can be computed using Higgs effective Lagrangian, where the top quark loop is integrated out \cite{Wilczek:1977zn, Shifman:1979eb, Dawson:1990zj, Djouadi:1991tka, Kniehl:1995tn, Chetyrkin:1997sg, Chetyrkin:1997un}
\begin{equation}
{\cal L}_{\rm eff} = \hat{C}_0 H \mathcal{O}_0 + {1\over m_{\rm t}^2} \sum_{i=1}^4 \hat{C}_i H \mathcal{O}_i + {\cal O}\left( {1\over m_{\rm t}^4} \right) \,.
\label{eq:HiggsEFT}
\end{equation}
Equivalently, the Higgs amplitudes can be understood as form factors, which are matrix elements of a local operator ${\cal O}_i$ and $n$ on-shell partons
\begin{equation}
{\cal F}_{{\cal O}_i,n} = \int {\rm d}^4 x \, {\rm e}^{-{\rm i} q\cdot x} \langle p_1, \ldots, p_n | {\cal O}_i(x) |0 \rangle \,.
\end{equation}
The operator ${\cal O}_i$ corresponds to a Higgs-gluon interaction vertex $H{\cal O}_i$ in the Higgs effective Lagrangian \eqref{eq:HiggsEFT}.
The leading terms contain dimension-4 and 6 operators \cite{Buchmuller:1985jz, Gracey:2002he, Neill:2009tn, Harlander:2013oja, Dawson:2014ora}
\begin{align}
{\cal O}_0 & =  {\rm tr}(F_{\mu\nu}F^{\mu\nu}) \,, \\ 
{\cal O}_{1} & = {\rm tr}(F_\mu^{~\nu} F_\nu^{~\rho} F_\rho^{~\mu}) \,, \\ 
{\cal O}_{2} & = {\rm tr}(D_\rho F_{\mu\nu} D^\rho F^{\mu\nu} ) \,, \\ 
{\cal O}_{3} & = {\rm tr}(D^\rho F_{\rho\mu} D_\sigma F^{\sigma\mu}) \,,\\
{\cal O}_{4} & = {\rm tr}(F_{\mu\rho} D^\rho D_\sigma F^{\sigma\mu}) \,.
\end{align}

To classify the operators, we introduce the \emph{length} of a given operator ${\cal O}$ as $L({\cal O})$, and it can be defined together with the \emph{minimal} form factor ${\cal F}_{{\cal O}, L({\cal O})}$, such that at tree-level, ${\cal F}^{(0)}_{{\cal O}, L({\cal O})}\neq0$ while ${\cal F}^{(0)}_{{\cal O}, n}=0$ when $n<L({\cal O})$.
For instance, the minimal form factor of ${\cal O}_0$ has two external gluons, so the length of ${\cal O}_0$ is two. Similarly, ${\cal O}_1$ has length three.
The lengths of ${\cal O}_{4}$ and  ${\cal O}_{3}$ are more subtle and require some explanation. Using the equation of motion $D^\mu F_{\mu\nu} \sim \sum_{i=1}^{n_f}(\bar\psi_i\gamma^\nu T^a \psi_i)$ (where $n_f$ is the flavor number), the operator ${\cal O}_{4}$ and  ${\cal O}_4' = F_{\mu\nu}^a D^\mu \sum_{i=1}^{n_f}(\bar\psi_i\gamma^\nu T^a \psi_i)$ are actually equivalent by equation of motion. This means that the minimal form factor of ${\cal O}_4$ requires two external quarks and one gluon, thus the length of ${\cal O}_4$ is three. Similar argument shows that the length of  ${\cal O}_3$ is four.

We can further classify operators according to their color structures, for which we introduce a diagrammatic notation: 
the blob {\includegraphics[height=.4cm]{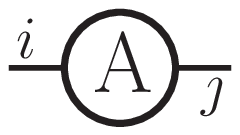}} represents an adjoint field and {\includegraphics[height=.4cm]{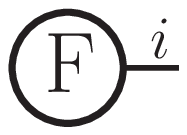}} represents a fundamental field, and $i,j$ are fundamental or anti-fundamental color indices. 
By multiplying several fields (blobs) and contracting the color indices, one obtains a color-singlet operator. 
The number of blobs matches the length of the operator. For example:
\begin{align}
\textrm{Length-2:} \ & \begin{tabular}{c}{\includegraphics[height=.54cm]{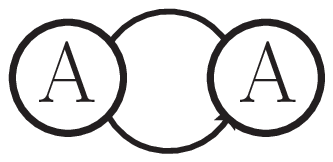}}\end{tabular}: \ {\rm tr}(F^2) \, , \quad \begin{tabular}{c}{\includegraphics[height=.5cm]{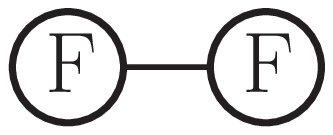}}\end{tabular}: \ \bar\psi\psi \,, \bar\phi\phi \,; \nonumber\\
\textrm{Length-3:} \ & \begin{tabular}{c}{\includegraphics[height=.48cm]{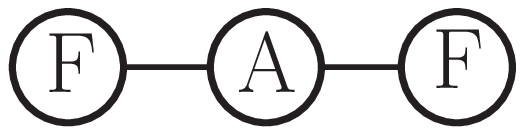}}\end{tabular}: \ F_{\mu\nu}^a(\bar\psi\gamma^{\mu\nu}T^a\psi) \,,
\label{eq:oper-examples}
\end{align}
where $\phi$ represent fundamental scalars in the scalar-YM theory (which is a cousin of QCD and can be obtained from the latter by replacing fundamental fermions to fundamental scalars).

Our study will focus on the form factors with three on-shell partons up to two loops.
Here we briefly explain our computation strategy, and we refer interested reader to \cite{Jin:2018fak} for detailed discussions.
The form factors with pure gluon external states can be obtained efficiently using unitarity based method \cite{Bern:1994zx, Bern:1994cg, Britto:2004nc} combined together with integration by parts (IBP) reduction \cite{Chetyrkin:1981qh, Tkachov:1981wb}. 
The form factors with external quark states have more complicated color structures, in particular the non-planar-color diagrams contribute.
We compute them using Feynman diagrams method with FeynArts \cite{Hahn:2000kx}.
In both computations, we convert tensor integrals to scalar integrals using the gauge invaraint basis, see e.g. \cite{Gehrmann:2011aa, Boels:2017gyc, Boels:2018nrr, Jin:2018fak}. 
IBP reduction  can then be applied to reduce the integrands to master integrals (e.g. with public codes \cite{Smirnov:2014hma, Lee:2013mka, vonManteuffel:2012np, Maierhoefer:2017hyi}). 
All master integrals we consider are known explicitly in terms of two-dimensional harmonic polylogarithms \cite{Gehrmann:2000zt,Gehrmann:2001ck, Gehrmann:2001jv}.

\section{Form factors and finite remainders}

Bare form factors contain ultraviolet (UV) and infrared (IR) divergences. We apply dimensional regularization ($D=4-2\epsilon$) in the conventional dimension regularization (CDR) scheme. The UV divergences come from both the coupling constant and the local operator, for which we apply the modified minimal subtraction renormalization ($\overline{\rm MS}$) scheme \cite{Bardeen:1978yd}. 
The renormalized form factors contain only IR divergences which take universal  Catani form as \cite{Catani:1998bh}: 
\begin{align}
{\cal F}_{\cal O}^{(1)} &= I^{(1)}(\epsilon) {\cal F}_{\cal O}^{(0)} + {\cal F}_{\cal O}^{(1),{\rm fin}} + {\cal O}(\epsilon) \,,  \\
{\cal F}_{\cal O}^{(2)} &= I^{(2)}(\epsilon) {\cal F}_{\cal O}^{(0)} +  I^{(1)}(\epsilon) {\cal F}_{\cal O}^{(1)} + {\cal F}_{\cal O}^{(2),{\rm fin}} + {\cal O}(\epsilon)   \,,
\label{eq:F2loopIRdecomp}
\end{align}
where $I^{(l)}$ are functions independent of operators. We describe the divergence subtractions explicitly in the Supplemental Material \cite{supplemental}. 

We have performed several non-trivial checks for our results. First, we reproduce the known results of form factors of ${\cal O}_0$ plus three partons \cite{Gehrmann:2011aa}. For $H\rightarrow 3g$, we apply both Feynman diagram and unitarity methods and find complete consistency. 
Second, our form factor results of ${\cal O}_i, i=0,1,2,4$ (which are computed independently) satisfy the required non-trivial linear relation \cite{Gracey:2002he}
\begin{align}
\label{eq:O2-linear-relation}
{\cal F}_{{\cal O}_{2}} = {1\over2}\, q^2 \,{\cal F}_{{\cal O}_{0}} -4\, g \, {\cal F}_{{\cal O}_{1}} +2\, {\cal F}_{{\cal O}_{4}} \,.
\end{align}
Third, our results reproduce the correct IR and UV divergences. 

The intrinsic new information of a form factor is contained in its finite part ${\cal F}_{\cal O}^{(l),{\rm fin}}$, which will be called the remainder function. We normalize the remainder by tree factor as $R_{\cal O}^{(l)} = {\cal F}_{\cal O}^{(l),{\rm fin}}/{\cal F}_{\cal O}^{(0)}$. Since the one-loop part is relatively simple, below we will focus on the two-loop form factors. At two loops, we can decompose the remainder function according to the general color structure as:
\begin{align}
\label{eq:fullRemainder}
{\cal R}_{{\cal O}}^{(2)}
= & \, C_A^2\, {\cal R}_{{\cal O}}^{(2),C_A^2} + C_A C_F {\cal R}_{{\cal O}}^{(2),C_A C_F} + C_F^2 {\cal R}_{{\cal O}}^{(2),C_F^2} \nonumber \\ 
+ & {n_f C_A}\, {\cal R}_{{\cal O}}^{(2),n_f C_A} + n_f C_F\, {\cal R}_{{\cal O}}^{(2),n_f C_F} + n_f^2\, {\cal R}_{{\cal O}}^{(2),n_f^2}  , 
\end{align}
where $C_A, C_F$ are the quadratic Casimirs in the adjoint and fundamental representations:
\begin{align}
C_A = N_c \,, \qquad C_F = \frac{N_c^2 -1}{2N_c} \,.
\label{eq:CACFinNc}
\end{align}

The maximal transcendentality degree of the two-loop remainder function is 4, and we can decompose it according to transcendentality degree as
\begin{align}
{\cal R}_{{\cal O}}^{(2)} = \sum_{d=0}^4 {\cal R}^{(2)}_{{\cal O}; d} ,
\end{align}
where ${\cal R}^{(2)}_{{\cal O};d}$ has uniform transcendentality degree $d$. 
We note that $n_f$ terms in \eqref{eq:fullRemainder} never appear in the maximal transcendentality (i.e.~degree-4) part.  Thus we have
\begin{equation}
\label{eq:deg4sum}
{\cal R}_{{\cal O};4}^{(2)}
= C_A^2\, {\cal R}_{{\cal O};4}^{(2),C_A^2} + C_A C_F {\cal R}_{{\cal O};4}^{(2),C_A C_F} + C_F^2 {\cal R}_{{\cal O};4}^{(2),C_F^2} .
\end{equation}

In the next two sections, we will focus on the maximal transcendentality parts and show that all the form factors we consider satisfy the maximal transcendentality principle. 

\section{Maximal Transcendentality Principle: Length-2 Cases}

Let us first consider the QCD form factors of the length-2 operator ${\cal O}_0$=${\rm tr}(F^2)$, which were first computed in \cite{Gehrmann:2011aa}. 

For the form factor with three external gluons, ${\cal O}_0\rightarrow 3g$, the $C_A C_F$ and $C_F^2$ terms  are both zero, and the maximally transcendental part comes only from ${\cal R}_{{\cal O}_0;4}^{(2),C_A^2}$. It has been noted in \cite{Brandhuber:2012vm} that the remainders of the ${\cal O}_0\rightarrow 3g$ form factors satisfy the maximal transcendentality principle:
\begin{align}
 {\cal R}^{(2)}_{{\cal O}_0;4}(1^-, 2^-, 3^\pm) = {\cal R}^{(2), {\cal N}=4}_{{\cal O}_0;4} = C_A^2 R^{(2)}_{\textrm{len-2};4} \,.
\end{align}
The same function was also obtained in the non-BPS Konishi form factor in ${\cal N}=4$ SYM \cite{Banerjee:2016kri}.

Above we have introduced the function $R^{(2)}_{\textrm{len-2};4}$, where the subscript `len-2' refers to length-2.
As we will see below, this function is a universal function for the form factors of length-2 operators. 
Its explicit expression can be given as:
\begin{align}
& R^{(2)}_{\textrm{len-2};4} = 
-2 \left[ J_4 \left( -\frac{u v}{w}\right)+J_4 \left( -\frac{v w}{u}\right)+J_4 \left( -\frac{w u}{v}\right)\right] 
\nonumber\\
& -8 \sum_{i=1}^3 \left[ \mathrm{Li}_4 \Big(1-\frac{1}{u_i}\Big)+\frac{\log^4 u_i}{4!} \right] -2 \left[ \sum_{i=1}^3 \mathrm{Li}_2 \Big(1-\frac{1}{u_i}\Big) \right]^2
\nonumber\\
& +\frac{1}{2} \left[ \sum_{i=1}^3 \log^2 u_i\right]^2 + 2 (J_2^2 - \zeta_2 J_2)  -\frac{\log^4(u v w)}{4!} \nonumber\\
& - \zeta_3 {\log(u v w)} - \frac{123}{8} \zeta_4   ,
\label{eq:RL2}
\end{align}
where
\begin{align}
& J_4(x)  =  \, \mathrm{Li}_4(x)-\log(-x) \mathrm{Li}_3(x)+\frac{\log^2(-x)}{2!} \mathrm{Li}_2(x) \nonumber\\
& \qquad\qquad -\frac{\log^3(-x)}{3!} \mathrm{Li}_1(x) - \frac{\log^4(-x)}{48} \,, \\
& J_2  =  \, \sum_{i=1}^3 \bigg( \mathrm{Li}_2(1-u_i)+{1\over2}\log(u_i)\log(u_{i+1}) \bigg) \,,
\end{align}
and 
\begin{align}
u = u_1 = {s_{12} \over s_{123} } \,, \ \  v = u_2 = {s_{23} \over s_{123}} \,, \ \  w = u_3 = {s_{13} \over s_{123}} \, .
\end{align}
We point out that the above result is computed with the Catani IR subtraction and is different from the ${\cal N}=4$ result in \cite{Brandhuber:2012vm} using the BDS subtraction \cite{Bern:2005iz} (see also \cite{Duhr:2012fh}). The difference is only from the change of subtraction schemes, and we have checked that when using Catani subtraction scheme, the ${\cal N}=4$ remainder is indeed equivalent to \eqref{eq:RL2}. 

The more interesting case is the form factor with quark external states: ${\cal O}_0\rightarrow q\bar{q}g$. 
In this case, both the $C_A C_F$ and $C_F^2$ terms in \eqref{eq:deg4sum} have non-trivial transcendentality degree-4 contributions. (Their explicit expressions are given in the Supplemental Material \cite{supplemental}.)
Remarkably, the direct sum of three terms in \eqref{eq:deg4sum} reproduces precisely the gluon remainder
\begin{align}
& {\cal R}^{(2), C_A^2}_{{\cal O}_0;4}(1^q, 2^{\bar q}, 3^\pm) + {\cal R}^{(2),C_A C_F}_{{\cal O}_0;4}(1^q, 2^{\bar q}, 3^\pm) \nonumber\\
& +  {\cal R}^{(2), C_F^2}_{{\cal O}_0;4}(1^q, 2^{\bar q}, 3^\pm) = R^{(2)}_{\textrm{len-2};4}   \,.
\label{eq:L2-quarksFF-relation}
\end{align}
Comparing with  \eqref{eq:deg4sum}, one can note that above sum makes sense if one makes a replacement for the color factor as
 \begin{equation}
 C_F \ \rightarrow \ C_A \,,
\end{equation}
so that all three terms share the same color factor $C_A^2$.
Such an identification for color factors has a  natural physical interpretation as changing the fermion representation from fundamental to adjoint. 
Similar relation was previously known for kinematic independent quantities such as anomalous dimensions \cite{Kotikov:2002ab, Kotikov:2004er,Dixon:2017nat}.\footnote{The universal maximally transcendental function was also noted for certain pseudo-scalar Higgs amplitudes involving $q\bar{q}g$ states \cite{Banerjee:2017faz}.}

We would like to stress that the above relation is rather non-trivial. 
First, the ${\cal O}_0\rightarrow 3g$ and ${\cal O}_0\rightarrow q\bar{q}g$  results are very different from each other. 
In particular, unlike the 3-gluon case, the latter have non-zero $C_A C_F$ and $C_F^2$ parts, and  both of them contain non-trivial 2d Harmonic polylogarithms of degree-4 (see the Supplemental Material \cite{supplemental}).  Furthermore, the ${\cal O}_0\rightarrow 3g$ case enjoys a permutational symmetry, while in ${\cal O}_0\rightarrow q\bar{q}g$ only a flip symmetry $(v\leftrightarrow w)$ is left.

In order to see if this relation applies to more general cases, we also compute the form factor of the length-2 operator, $\bar\psi \psi$. As shown in \eqref{eq:oper-examples}, this operator has very different color structure compared to ${\cal O}_0$.
It turns out that its maximally transcendental part satisfies the same relation: 
\begin{align}
& {\cal R}^{(2), C_A^2}_{\bar\psi \psi;4}(1^q, 2^{\bar q}, 3^\pm) + {\cal R}^{(2),C_A C_F}_{\bar\psi \psi;4}(1^q, 2^{\bar q}, 3^\pm) \nonumber\\
& +  {\cal R}^{(2), C_F^2}_{\bar\psi \psi;4}(1^q, 2^{\bar q}, 3^\pm)  = R^{(2)}_{\textrm{len-2};4}\,.
\label{eq:L2-psipsi-relation}
\end{align}
The explicit expressions of three terms are given in the Supplemental Material \cite{supplemental}.

We have also considered form factors of operator $\bar\phi\phi$ in the scalar-YM theory. It turns out that its maximally transcendental part is exactly the same as the $\bar\psi \psi$ result in QCD, without changing any color factors. 
The equivalence between fermion and scalar cases implies that the correspondence does not depend on the spin of the fields. 

Finally, we note that each terms in \eqref{eq:L2-psipsi-relation} are different from those in \eqref{eq:L2-quarksFF-relation}. 
This difference is not surprising, since the operators ${\rm tr}(F^2)$ and  $\{\bar\psi\psi, \bar\phi\phi\}$ have different color structures, as indicated in \eqref{eq:oper-examples}.

\section{Maximal Transcendentality Principle: Length-3 Cases}

In this section, we consider further the form factors with length-3 operators. As we will see, despite that the expressions are very different between the length-2 and length-3 cases,  the maximal transcendentality principle still holds. 

We first consider the form factors of ${\cal O}_1$=${\rm tr}(F^3)$ with three external gluon states.  The $C_A C_F$ and $C_F^2$ terms are both zero. It has been observed that the maximally transcendental part of the ${\cal O}_1\rightarrow 3g$ form factor remainders are the same in QCD and ${\cal N}=4$ SYM \cite{Brandhuber:2017bkg, Jin:2018fak}:
\begin{equation}
{\cal R}^{(2)}_{{\cal O}_1;4}(1^-, 2^-, 3^-) = {\cal R}^{(2), {\cal N}=4}_{{\cal O}_1;4} = C_A^2 R^{(2)}_{\textrm{len-3};4}   \,,
\end{equation}
where we introduce $R^{(2)}_{\textrm{len-3};4}$.
Its explicit form can be given as \cite{Jin:2018fak}
\begin{align}
& R^{(2)}_{\textrm{len-3};4} =  
-{3\over2} {\rm Li}_4(u) + {3\over4} {\rm Li}_4\left(-{u v \over w} \right) - {3\over2} \log(w)  {\rm Li}_3 \left(-{u\over v} \right) \nonumber\\
& + {\log^2(u) \over 32} \left[ \log^2(u) + 2\log^2(v) - 4\log(v)\log(w) \right]  
\label{eq:Rlen-3} \\
& + {\zeta_2 \over 8} \left[ 5\log^2(u) - 2 \log(v)\log(w) \right]- {1\over4} \zeta_4 
+ \textrm{perms}(u,v,w) . \nonumber
\end{align}
We have also computed the $H\rightarrow 3g$ amplitudes with higher dimension length-3 operators in the pure gluon sector of Higgs effective Lagrangian, and they all share the the same maximally transcendental part.

To study the form factors with external quarks, we consider the length-3 operator ${\cal O}_4\sim F_{\mu\nu}D^\mu(\bar\psi\gamma^\nu\psi)$. In this case, the $C_A C_F$ and $C_F^2$ terms of ${\cal O}_4\rightarrow q \bar{q}g$ have non-trivial contributions.
Remarkably, they satisfy the same correspondence as in the length-2 cases: by changing $C_F$ to be $C_A$, the maximally transcendental part are identical to the 3-gluon case:
\begin{align}
& {\cal R}^{(2)}_{{\cal O}_4;4}(1^q, 2^{\bar q}, 3^\pm) \big|_{C_F \rightarrow C_A} \label{eq:FpsipsiQQGSum} \\
& = C_A^2 \Big( {\cal R}^{(2), C_A^2}_{{\cal O}_4;4} + {\cal R}^{(2), C_A C_F}_{{\cal O}_4;4} + {\cal R}^{(2), C_F^2}_{{\cal O}_4;4} \Big) = C_A^2 R^{(2)}_{\textrm{len-3};4}  \,.  \nonumber
\end{align}
The explicit expressions of the three terms with different color factors are given in terms of 2d Harmonic polylogarithms in the Supplemental Material \cite{supplemental}.

\begin{figure}[tt]
\centering
\includegraphics[clip,scale=0.32]{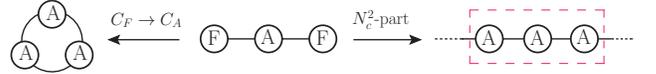}
\caption[a]{The correspondences of form factors with length-3 operators. The red square on the right indicates a density contribution in the higher length case.}
\label{fig:FAFvsAAA}
\end{figure}

Let us mention another interesting relation. If we reorganize the remainder of ${\cal O}_4\rightarrow q \bar{q}g$ in terms of $N_c$ expansion using \eqref{eq:CACFinNc},
the term with color factor $N_c^2$ is
\begin{align}
\label{eq:FpsipsiQQGNc2}
& {\cal R}^{(2), N_c^2}_{{\cal O}_4;4}(1^q, 2^{\bar q}, 3^\pm) =   G(1-v,1-v,1,0,w)
\\
& -\text{Li}_4(1-v)-\text{Li}_4(v) + \text{Li}_4\Big(\frac{v-1}{v}\Big)  \nonumber \\
&
+\text{Li}_3(v) \log \Big(\frac{u}{1-v}\Big)+ \text{Li}_3\Big(\frac{u}{1-v}\Big)\log (v) 
\nonumber\\
& + \big[ \text{Li}_3(v)+\text{Li}_3(1-v) \big] \left[ \log (w)-2 \log \Big(\frac{u}{1-v}\Big)\right]
\nonumber   \\ 
& + \text{Li}_2\Big(\frac{u}{1-v}\Big) \text{Li}_2\Big(\frac{v-1}{v}\Big) +\frac{1}{2} \text{Li}_2(v) \log
   ^2\Big(\frac{u}{1-v}\Big)
\nonumber\\
&+\text{Li}_2(1-v) \log (1-v) \log \Big(\frac{u}{1-v}\Big)
+ \frac{1}{24} \log ^2(v) \nonumber \\ 
& \times \left[ \log (v) \log \Big(\frac{v}{(1-v)^4}\Big)-3 \log (w) \log
   \Big(\frac{w}{(1-v)^4}\Big) \right]
\nonumber  \\
& + \zeta_2 \left[ \text{Li}_2(1-v) +\log (v) \log \left(\frac{v}{w}\right) 
- {1\over2} \log ^2\Big(\frac{u}{1-v}\Big) \right] 
\nonumber\\
& +\zeta_3 \left[ \log \Big(\frac{u}{1-v}\Big)-5 \log (v)\right]   +\frac{23 \zeta_4}{8}
 +  \{v \, \leftrightarrow\, w\} \,, \nonumber
\end{align}
where we have simplified the expression using the symbol technique \cite{Goncharov:2010jf}. Strikingly, the symbol of this function is  identical to the universal partial density remainder of minimal form factors with higher length operators in ${\cal N}=4$ SYM, which was obtained first for ${\rm tr}(\phi^L)$ \cite{Brandhuber:2014ica} and later for more general operators \cite{Loebbert:2015ova, Brandhuber:2016fni, Loebbert:2016xkw}. Their functional forms only differ by very simple terms:
\begin{align}
& {\cal R}^{(2), N_c^2}_{{\cal O}_4;4}(1^q, 2^{\bar q}, 3^\pm) - {\cal R}^{(2), {\cal N}=4}_{\textrm{density};4}  = \frac{19}{4}\zeta_4-4 \zeta_3 \log(v w)  \,.
\end{align}
This coincidence is surprising in the sense that the remainder density in ${\cal N}=4$ SYM is a partial quantity for higher length minimal form factors, while \eqref{eq:FpsipsiQQGNc2} is the leading  $N_c$ color result for a length-3 form factor with fundamental quarks. 
This suggests that the density for higher length form factors in QCD is independent of the representation as well as the spin of the particles. 
The leading $N_c$ contribution for the minimal form factor with a length-$L$ operator, ${\cal O}_L \in \begin{tabular}{c}{\includegraphics[height=.47cm]{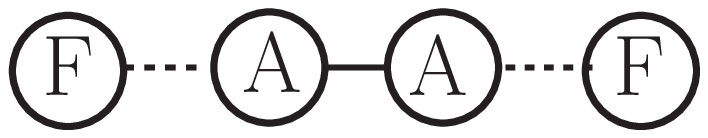}}\end{tabular}$, is expected to be given by:
\begin{equation}
{\cal R}^{(2), N_c^2}_{{\cal O}_L;4}(1^{\bar q}, 2^g, \ldots, (L-1)^g, L^q) \simeq \sum_{i=1}^{L-2} {\cal R}^{(2)}_{\textrm{density};4}(u_i,v_i,w_i) \,,
\end{equation}
up to simple terms containing $\zeta_3, \zeta_4$.
We summarize the above correspondences in Figure.\,\ref{fig:FAFvsAAA}.

For completeness, we give the result with $N_c^{-2}$ color factor:
\begin{align}
\label{eq:FpsipsiQQGNcm2}
{\cal R}^{(2), 1/N_c^2}_{{\cal O}_4;4}(1^q, 2^{\bar q}, 3^\pm) =  - \frac{11}{2}\zeta_4 + 6 \zeta_3 \log(u) \,.
\end{align}
The $N_c^0$ part can be obtained using (\ref{eq:FpsipsiQQGSum}), (\ref{eq:FpsipsiQQGNc2}), and (\ref{eq:FpsipsiQQGNcm2}). 

Finally, let us mention that in the scalar-YM theory, the form factor with length-3 operator $F_{\mu\nu}^a D^\mu \sum_{i=1}^{n_f}(\bar\phi_i  \overleftrightarrow{D^\mu}T^a \phi_i)$ (which is a scalar version of  ${\cal O}_{4}$) have identical maximally transcendental parts as the QCD results without changing any color factors, similar to the length-2 case.

\begin{table*}[t]
  \begin{center}
  \caption{The universal maximally transcendental properties for Higgs amplitudes or form factors of length-2 and 3 operators with three partons, and minimal form factors with higher length operators are summarized. The color-singlet operators are classified according to their lengths and representative examples are provided. We also indicate the external on-shell partons.}
    \label{tab:summary}
    \begin{tabular}{c|c|c|c|c|c|c|c}
    \hline\hline
     & \multicolumn{3}{c|}{\textbf{Length-2}} &  \multicolumn{2}{c|}{\textbf{Length-3}} & \multicolumn{2}{c}{\textbf{Higher length}} \\ \hline
     Operators & \begin{tabular}{c}{\includegraphics[height=.5cm]{AA}}\end{tabular} & 
     \multicolumn{2}{c|}{\begin{tabular}{c}{\includegraphics[height=.5cm]{FF}}\end{tabular}} & 
     \begin{tabular}{c}{\includegraphics[height=.9cm]{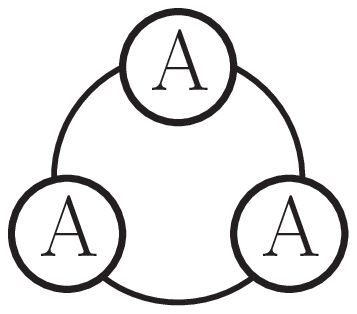}}\end{tabular} & 
     \begin{tabular}{c}{\includegraphics[height=.5cm]{FAF}}\end{tabular} & 
     \begin{tabular}{c}{\includegraphics[height=.56cm]{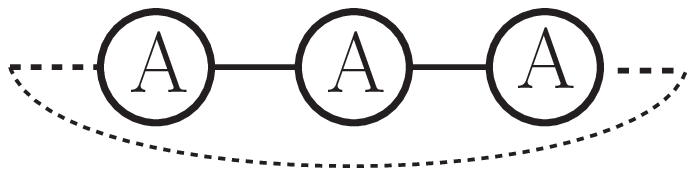}}\end{tabular}& 
     \begin{tabular}{c}{\includegraphics[height=.47cm]{FAAFdensity}}\end{tabular} \\\hline
     Examples & ${\rm tr}(F^2)$ & $\bar\psi\psi$ & $\bar\phi  \phi$ & 
     $\begin{matrix} {\rm tr}(F^3), \\ {\rm tr}(F_\mu^{\,\nu} D_\sigma F_\nu^{\,\rho} D^\sigma F_\rho^{\,\mu}) \end{matrix}$ & 
     $\begin{matrix} F_{\mu\nu}D^\mu(\bar\psi\gamma^\nu\psi), \\ F_{\mu\nu}(\bar\psi\gamma^{\mu\nu}\psi) \end{matrix}$ & 
     ${\rm tr}(F^L), L\geq4$ & $ \bar\psi(F^L)\psi, L\geq2$ \\\hline
     External Partons & $(g,g,g), (\bar\psi,\psi,g)$ & $(\bar\psi,\psi,g)$ & $(\bar\phi,\phi,g)$ & $(g,g,g)$ & $(\bar\psi,g, \psi)$ & $(g_1, \ldots, g_L)$ & $(\bar\psi,g_1, \ldots, g_L,\psi)$ \\\hline
     \begin{tabular}{c} Max. Trans. \\ \textrm{Remainder} \\ (\textrm{with }{$C_F\rightarrow C_A$}) \end{tabular} 
     & \multicolumn{3}{c|}{$R_{\rm L2;4}(u,v,w)$} 
     & \multicolumn{2}{c|}{$R_{\rm len-3;4}(u,v,w)$} 
     & \multicolumn{2}{c}{$\sum_i {\cal R}^{(2)}_{\textrm{density};4}(u_i,v_i,w_i)$} \\ \hline\hline
    \end{tabular}
  \end{center}
\end{table*}

\section{Summary and Discussion}

In this paper, we generalize the maximal transcendentality principle to Higgs amplitudes and form factors that contain external fundamental quarks in QCD. By a simple change of color factors, the maximally transcendental parts of the $H \rightarrow q \bar q g$ amplitudes become identical to the $H \rightarrow 3g$ amplitudes. 
The correspondence is found to be true for both length-2 and length-3 operators. The universal maximally transcendental parts are given in \eqref{eq:RL2} and \eqref{eq:Rlen-3}, for the length-2 and length-3 cases respectively. We also find that the leading $N_c$ term of $H \rightarrow q \bar q g$ amplitudes with length-3 operator is equivalent to the ${\cal N}=4$ remainder density for higher length operators. A number of non-trivial new results have been obtained to test the correspondence, including Higgs amplitudes with higher dimensional operators in the effective Lagrangian, and also form factors in both QCD and scalar-YM theory. 
We summarize the correspondence in Table~\ref{tab:summary}.

Let us comment on a few open problems regarding the correspondence. First, the maximal transcendentality principle allows one to obtain the maximal transcendentality part (functionally the most complicated part) in Higgs amplitudes from their ${\cal N}=4$ counterparts. 
It would be interesting to check if the relation holds for more general cases, such as at three loops, or the Higgs plus four-parton amplitudes \cite{Dixon:2009uk, Badger:2009hw, Badger:2009vh}.
Second, the knowledge of lower transcendentality parts are also important, in order to obtain full QCD results.
Some evidence fo the relations of the lower transcendental parts was found in \cite{Jin:2018fak}. It would be interesting to explore this further.
Third, via unitarity cuts, gluon/quark amplitudes (without the Higgs particle) are building blocks of form factors.
The correspondence we found indicates that there could be hidden relations (induced by unitarity cuts) for those amplitudes.
Finally, the universal relations and the simplicity of the results we present in this paper are hard to understand using standard Feynman diagram methods. A solid understanding of their origin is expected to lead to a better way of computing amplitudes or form factors. 
We hope to explore these in the future work.

\section{Acknowledgement}
We thank Hui Luo, Jianping Ma, and Ke Ren for interesting discussions.
This work is supported in part by the National Natural Science Foundation of China (Grants No.~11822508, 11947302, 11935013),
by the Chinese Academy of Sciences (CAS) Hundred-Talent Program, 
and by the Key Research Program of Frontier Sciences of CAS. 
We also thank the support of the HPC Cluster of ITP-CAS.

\bibliographystyle{apsrev4-1}

\begin{thebibliography}{73}%
\makeatletter
\providecommand \@ifxundefined [1]{%
 \@ifx{#1\undefined}
}%
\providecommand \@ifnum [1]{%
 \ifnum #1\expandafter \@firstoftwo
 \else \expandafter \@secondoftwo
 \fi
}%
\providecommand \@ifx [1]{%
 \ifx #1\expandafter \@firstoftwo
 \else \expandafter \@secondoftwo
 \fi
}%
\providecommand \natexlab [1]{#1}%
\providecommand \enquote  [1]{``#1''}%
\providecommand \bibnamefont  [1]{#1}%
\providecommand \bibfnamefont [1]{#1}%
\providecommand \citenamefont [1]{#1}%
\providecommand \href@noop [0]{\@secondoftwo}%
\providecommand \href [0]{\begingroup \@sanitize@url \@href}%
\providecommand \@href[1]{\@@startlink{#1}\@@href}%
\providecommand \@@href[1]{\endgroup#1\@@endlink}%
\providecommand \@sanitize@url [0]{\catcode `\\12\catcode `\$12\catcode
  `\&12\catcode `\#12\catcode `\^12\catcode `\_12\catcode `\%12\relax}%
\providecommand \@@startlink[1]{}%
\providecommand \@@endlink[0]{}%
\providecommand \url  [0]{\begingroup\@sanitize@url \@url }%
\providecommand \@url [1]{\endgroup\@href {#1}{\urlprefix }}%
\providecommand \urlprefix  [0]{URL }%
\providecommand \Eprint [0]{\href }%
\providecommand \doibase [0]{http://dx.doi.org/}%
\providecommand \selectlanguage [0]{\@gobble}%
\providecommand \bibinfo  [0]{\@secondoftwo}%
\providecommand \bibfield  [0]{\@secondoftwo}%
\providecommand \translation [1]{[#1]}%
\providecommand \BibitemOpen [0]{}%
\providecommand \bibitemStop [0]{}%
\providecommand \bibitemNoStop [0]{.\EOS\space}%
\providecommand \EOS [0]{\spacefactor3000\relax}%
\providecommand \BibitemShut  [1]{\csname bibitem#1\endcsname}%
\let\auto@bib@innerbib\@empty
\bibitem [{\citenamefont {Parke}\ and\ \citenamefont
  {Taylor}(1986)}]{Parke:1986gb}%
  \BibitemOpen
  \bibfield  {author} {\bibinfo {author} {\bibfnamefont {S.~J.}\ \bibnamefont
  {Parke}}\ and\ \bibinfo {author} {\bibfnamefont {T.~R.}\ \bibnamefont
  {Taylor}},\ }\href {\doibase 10.1103/PhysRevLett.56.2459} {\bibfield
  {journal} {\bibinfo  {journal} {Phys. Rev. Lett.}\ }\textbf {\bibinfo
  {volume} {56}},\ \bibinfo {pages} {2459} (\bibinfo {year}
  {1986})}\BibitemShut {NoStop}%
\bibitem [{\citenamefont {Del~Duca}\ \emph {et~al.}(2010)\citenamefont
  {Del~Duca}, \citenamefont {Duhr},\ and\ \citenamefont
  {Smirnov}}]{DelDuca:2010zg}%
  \BibitemOpen
  \bibfield  {author} {\bibinfo {author} {\bibfnamefont {V.}~\bibnamefont
  {Del~Duca}}, \bibinfo {author} {\bibfnamefont {C.}~\bibnamefont {Duhr}}, \
  and\ \bibinfo {author} {\bibfnamefont {V.~A.}\ \bibnamefont {Smirnov}},\
  }\href {\doibase 10.1007/JHEP05(2010)084} {\bibfield  {journal} {\bibinfo
  {journal} {JHEP}\ }\textbf {\bibinfo {volume} {05}},\ \bibinfo {pages} {084}
  (\bibinfo {year} {2010})},\ \Eprint {http://arxiv.org/abs/1003.1702}
  {arXiv:1003.1702 [hep-th]} \BibitemShut {NoStop}%
\bibitem [{\citenamefont {Goncharov}\ \emph {et~al.}(2010)\citenamefont
  {Goncharov}, \citenamefont {Spradlin}, \citenamefont {Vergu},\ and\
  \citenamefont {Volovich}}]{Goncharov:2010jf}%
  \BibitemOpen
  \bibfield  {author} {\bibinfo {author} {\bibfnamefont {A.~B.}\ \bibnamefont
  {Goncharov}}, \bibinfo {author} {\bibfnamefont {M.}~\bibnamefont {Spradlin}},
  \bibinfo {author} {\bibfnamefont {C.}~\bibnamefont {Vergu}}, \ and\ \bibinfo
  {author} {\bibfnamefont {A.}~\bibnamefont {Volovich}},\ }\href {\doibase
  10.1103/PhysRevLett.105.151605} {\bibfield  {journal} {\bibinfo  {journal}
  {Phys. Rev. Lett.}\ }\textbf {\bibinfo {volume} {105}},\ \bibinfo {pages}
  {151605} (\bibinfo {year} {2010})},\ \Eprint {http://arxiv.org/abs/1006.5703}
  {arXiv:1006.5703 [hep-th]} \BibitemShut {NoStop}%
\bibitem [{\citenamefont {Gehrmann}\ \emph {et~al.}(2016)\citenamefont
  {Gehrmann}, \citenamefont {Henn},\ and\ \citenamefont
  {Lo~Presti}}]{Gehrmann:2015bfy}%
  \BibitemOpen
  \bibfield  {author} {\bibinfo {author} {\bibfnamefont {T.}~\bibnamefont
  {Gehrmann}}, \bibinfo {author} {\bibfnamefont {J.~M.}\ \bibnamefont {Henn}},
  \ and\ \bibinfo {author} {\bibfnamefont {N.~A.}\ \bibnamefont {Lo~Presti}},\
  }\href {\doibase 10.1103/PhysRevLett.116.189903,
  10.1103/PhysRevLett.116.062001} {\bibfield  {journal} {\bibinfo  {journal}
  {Phys. Rev. Lett.}\ }\textbf {\bibinfo {volume} {116}},\ \bibinfo {pages}
  {062001} (\bibinfo {year} {2016})},\ \bibinfo {note} {[Erratum: Phys. Rev.
  Lett.116,no.18,189903(2016)]},\ \Eprint {http://arxiv.org/abs/1511.05409}
  {arXiv:1511.05409 [hep-ph]} \BibitemShut {NoStop}%
\bibitem [{\citenamefont {Dunbar}\ \emph {et~al.}(2016)\citenamefont {Dunbar},
  \citenamefont {Jehu},\ and\ \citenamefont {Perkins}}]{Dunbar:2016gjb}%
  \BibitemOpen
  \bibfield  {author} {\bibinfo {author} {\bibfnamefont {D.~C.}\ \bibnamefont
  {Dunbar}}, \bibinfo {author} {\bibfnamefont {G.~R.}\ \bibnamefont {Jehu}}, \
  and\ \bibinfo {author} {\bibfnamefont {W.~B.}\ \bibnamefont {Perkins}},\
  }\href {\doibase 10.1103/PhysRevLett.117.061602} {\bibfield  {journal}
  {\bibinfo  {journal} {Phys. Rev. Lett.}\ }\textbf {\bibinfo {volume} {117}},\
  \bibinfo {pages} {061602} (\bibinfo {year} {2016})},\ \Eprint
  {http://arxiv.org/abs/1605.06351} {arXiv:1605.06351 [hep-th]} \BibitemShut
  {NoStop}%
\bibitem [{\citenamefont {Dunbar}\ \emph {et~al.}(2017)\citenamefont {Dunbar},
  \citenamefont {Godwin}, \citenamefont {Jehu},\ and\ \citenamefont
  {Perkins}}]{Dunbar:2017nfy}%
  \BibitemOpen
  \bibfield  {author} {\bibinfo {author} {\bibfnamefont {D.~C.}\ \bibnamefont
  {Dunbar}}, \bibinfo {author} {\bibfnamefont {J.~H.}\ \bibnamefont {Godwin}},
  \bibinfo {author} {\bibfnamefont {G.~R.}\ \bibnamefont {Jehu}}, \ and\
  \bibinfo {author} {\bibfnamefont {W.~B.}\ \bibnamefont {Perkins}},\ }\href
  {\doibase 10.1103/PhysRevD.96.116013} {\bibfield  {journal} {\bibinfo
  {journal} {Phys. Rev.}\ }\textbf {\bibinfo {volume} {D96}},\ \bibinfo {pages}
  {116013} (\bibinfo {year} {2017})},\ \Eprint
  {http://arxiv.org/abs/1710.10071} {arXiv:1710.10071 [hep-th]} \BibitemShut
  {NoStop}%
\bibitem [{\citenamefont {Badger}\ \emph {et~al.}(2019)\citenamefont {Badger},
  \citenamefont {Brønnum-Hansen}, \citenamefont {Hartanto},\ and\
  \citenamefont {Peraro}}]{Badger:2018enw}%
  \BibitemOpen
  \bibfield  {author} {\bibinfo {author} {\bibfnamefont {S.}~\bibnamefont
  {Badger}}, \bibinfo {author} {\bibfnamefont {C.}~\bibnamefont
  {Brønnum-Hansen}}, \bibinfo {author} {\bibfnamefont {H.~B.}\ \bibnamefont
  {Hartanto}}, \ and\ \bibinfo {author} {\bibfnamefont {T.}~\bibnamefont
  {Peraro}},\ }\href {\doibase 10.1007/JHEP01(2019)186} {\bibfield  {journal}
  {\bibinfo  {journal} {JHEP}\ }\textbf {\bibinfo {volume} {01}},\ \bibinfo
  {pages} {186} (\bibinfo {year} {2019})},\ \Eprint
  {http://arxiv.org/abs/1811.11699} {arXiv:1811.11699 [hep-ph]} \BibitemShut
  {NoStop}%
\bibitem [{\citenamefont {Abreu}\ \emph
  {et~al.}(2019{\natexlab{a}})\citenamefont {Abreu}, \citenamefont {Dormans},
  \citenamefont {Febres~Cordero}, \citenamefont {Ita},\ and\ \citenamefont
  {Page}}]{Abreu:2018zmy}%
  \BibitemOpen
  \bibfield  {author} {\bibinfo {author} {\bibfnamefont {S.}~\bibnamefont
  {Abreu}}, \bibinfo {author} {\bibfnamefont {J.}~\bibnamefont {Dormans}},
  \bibinfo {author} {\bibfnamefont {F.}~\bibnamefont {Febres~Cordero}},
  \bibinfo {author} {\bibfnamefont {H.}~\bibnamefont {Ita}}, \ and\ \bibinfo
  {author} {\bibfnamefont {B.}~\bibnamefont {Page}},\ }\href {\doibase
  10.1103/PhysRevLett.122.082002} {\bibfield  {journal} {\bibinfo  {journal}
  {Phys. Rev. Lett.}\ }\textbf {\bibinfo {volume} {122}},\ \bibinfo {pages}
  {082002} (\bibinfo {year} {2019}{\natexlab{a}})},\ \Eprint
  {http://arxiv.org/abs/1812.04586} {arXiv:1812.04586 [hep-ph]} \BibitemShut
  {NoStop}%
\bibitem [{\citenamefont {Abreu}\ \emph
  {et~al.}(2019{\natexlab{b}})\citenamefont {Abreu}, \citenamefont {Dormans},
  \citenamefont {Febres~Cordero}, \citenamefont {Ita}, \citenamefont {Page},\
  and\ \citenamefont {Sotnikov}}]{Abreu:2019odu}%
  \BibitemOpen
  \bibfield  {author} {\bibinfo {author} {\bibfnamefont {S.}~\bibnamefont
  {Abreu}}, \bibinfo {author} {\bibfnamefont {J.}~\bibnamefont {Dormans}},
  \bibinfo {author} {\bibfnamefont {F.}~\bibnamefont {Febres~Cordero}},
  \bibinfo {author} {\bibfnamefont {H.}~\bibnamefont {Ita}}, \bibinfo {author}
  {\bibfnamefont {B.}~\bibnamefont {Page}}, \ and\ \bibinfo {author}
  {\bibfnamefont {V.}~\bibnamefont {Sotnikov}},\ }\href@noop {} {\  (\bibinfo
  {year} {2019}{\natexlab{b}})},\ \Eprint {http://arxiv.org/abs/1904.00945}
  {arXiv:1904.00945 [hep-ph]} \BibitemShut {NoStop}%
\bibitem [{\citenamefont {Wilczek}(1977)}]{Wilczek:1977zn}%
  \BibitemOpen
  \bibfield  {author} {\bibinfo {author} {\bibfnamefont {F.}~\bibnamefont
  {Wilczek}},\ }\href {\doibase 10.1103/PhysRevLett.39.1304} {\bibfield
  {journal} {\bibinfo  {journal} {Phys. Rev. Lett.}\ }\textbf {\bibinfo
  {volume} {39}},\ \bibinfo {pages} {1304} (\bibinfo {year}
  {1977})}\BibitemShut {NoStop}%
\bibitem [{\citenamefont {Shifman}\ \emph {et~al.}(1979)\citenamefont
  {Shifman}, \citenamefont {Vainshtein}, \citenamefont {Voloshin},\ and\
  \citenamefont {Zakharov}}]{Shifman:1979eb}%
  \BibitemOpen
  \bibfield  {author} {\bibinfo {author} {\bibfnamefont {M.~A.}\ \bibnamefont
  {Shifman}}, \bibinfo {author} {\bibfnamefont {A.~I.}\ \bibnamefont
  {Vainshtein}}, \bibinfo {author} {\bibfnamefont {M.~B.}\ \bibnamefont
  {Voloshin}}, \ and\ \bibinfo {author} {\bibfnamefont {V.~I.}\ \bibnamefont
  {Zakharov}},\ }\href@noop {} {\bibfield  {journal} {\bibinfo  {journal} {Sov.
  J. Nucl. Phys.}\ }\textbf {\bibinfo {volume} {30}},\ \bibinfo {pages} {711}
  (\bibinfo {year} {1979})},\ \bibinfo {note} {[Yad.
  Fiz.30,1368(1979)]}\BibitemShut {NoStop}%
\bibitem [{\citenamefont {Dawson}(1991)}]{Dawson:1990zj}%
  \BibitemOpen
  \bibfield  {author} {\bibinfo {author} {\bibfnamefont {S.}~\bibnamefont
  {Dawson}},\ }\href {\doibase 10.1016/0550-3213(91)90061-2} {\bibfield
  {journal} {\bibinfo  {journal} {Nucl. Phys.}\ }\textbf {\bibinfo {volume}
  {B359}},\ \bibinfo {pages} {283} (\bibinfo {year} {1991})}\BibitemShut
  {NoStop}%
\bibitem [{\citenamefont {Djouadi}\ \emph {et~al.}(1991)\citenamefont
  {Djouadi}, \citenamefont {Spira},\ and\ \citenamefont
  {Zerwas}}]{Djouadi:1991tka}%
  \BibitemOpen
  \bibfield  {author} {\bibinfo {author} {\bibfnamefont {A.}~\bibnamefont
  {Djouadi}}, \bibinfo {author} {\bibfnamefont {M.}~\bibnamefont {Spira}}, \
  and\ \bibinfo {author} {\bibfnamefont {P.~M.}\ \bibnamefont {Zerwas}},\
  }\href {\doibase 10.1016/0370-2693(91)90375-Z} {\bibfield  {journal}
  {\bibinfo  {journal} {Phys. Lett.}\ }\textbf {\bibinfo {volume} {B264}},\
  \bibinfo {pages} {440} (\bibinfo {year} {1991})}\BibitemShut {NoStop}%
\bibitem [{\citenamefont {Kniehl}\ and\ \citenamefont
  {Spira}(1995)}]{Kniehl:1995tn}%
  \BibitemOpen
  \bibfield  {author} {\bibinfo {author} {\bibfnamefont {B.~A.}\ \bibnamefont
  {Kniehl}}\ and\ \bibinfo {author} {\bibfnamefont {M.}~\bibnamefont {Spira}},\
  }\href {\doibase 10.1007/s002880050007} {\bibfield  {journal} {\bibinfo
  {journal} {Z. Phys.}\ }\textbf {\bibinfo {volume} {C69}},\ \bibinfo {pages}
  {77} (\bibinfo {year} {1995})},\ \Eprint
  {http://arxiv.org/abs/hep-ph/9505225} {arXiv:hep-ph/9505225 [hep-ph]}
  \BibitemShut {NoStop}%
\bibitem [{\citenamefont {Chetyrkin}\ \emph {et~al.}(1997)\citenamefont
  {Chetyrkin}, \citenamefont {Kniehl},\ and\ \citenamefont
  {Steinhauser}}]{Chetyrkin:1997sg}%
  \BibitemOpen
  \bibfield  {author} {\bibinfo {author} {\bibfnamefont {K.~G.}\ \bibnamefont
  {Chetyrkin}}, \bibinfo {author} {\bibfnamefont {B.~A.}\ \bibnamefont
  {Kniehl}}, \ and\ \bibinfo {author} {\bibfnamefont {M.}~\bibnamefont
  {Steinhauser}},\ }\href {\doibase 10.1103/PhysRevLett.79.2184} {\bibfield
  {journal} {\bibinfo  {journal} {Phys. Rev. Lett.}\ }\textbf {\bibinfo
  {volume} {79}},\ \bibinfo {pages} {2184} (\bibinfo {year} {1997})},\ \Eprint
  {http://arxiv.org/abs/hep-ph/9706430} {arXiv:hep-ph/9706430 [hep-ph]}
  \BibitemShut {NoStop}%
\bibitem [{\citenamefont {Chetyrkin}\ \emph {et~al.}(1998)\citenamefont
  {Chetyrkin}, \citenamefont {Kniehl},\ and\ \citenamefont
  {Steinhauser}}]{Chetyrkin:1997un}%
  \BibitemOpen
  \bibfield  {author} {\bibinfo {author} {\bibfnamefont {K.~G.}\ \bibnamefont
  {Chetyrkin}}, \bibinfo {author} {\bibfnamefont {B.~A.}\ \bibnamefont
  {Kniehl}}, \ and\ \bibinfo {author} {\bibfnamefont {M.}~\bibnamefont
  {Steinhauser}},\ }\href {\doibase 10.1016/S0550-3213(98)81004-3,
  10.1016/S0550-3213(97)00649-4} {\bibfield  {journal} {\bibinfo  {journal}
  {Nucl. Phys.}\ }\textbf {\bibinfo {volume} {B510}},\ \bibinfo {pages} {61}
  (\bibinfo {year} {1998})},\ \Eprint {http://arxiv.org/abs/hep-ph/9708255}
  {arXiv:hep-ph/9708255 [hep-ph]} \BibitemShut {NoStop}%
\bibitem [{\citenamefont {Boughezal}\ \emph {et~al.}(2013)\citenamefont
  {Boughezal}, \citenamefont {Caola}, \citenamefont {Melnikov}, \citenamefont
  {Petriello},\ and\ \citenamefont {Schulze}}]{Boughezal:2013uia}%
  \BibitemOpen
  \bibfield  {author} {\bibinfo {author} {\bibfnamefont {R.}~\bibnamefont
  {Boughezal}}, \bibinfo {author} {\bibfnamefont {F.}~\bibnamefont {Caola}},
  \bibinfo {author} {\bibfnamefont {K.}~\bibnamefont {Melnikov}}, \bibinfo
  {author} {\bibfnamefont {F.}~\bibnamefont {Petriello}}, \ and\ \bibinfo
  {author} {\bibfnamefont {M.}~\bibnamefont {Schulze}},\ }\href {\doibase
  10.1007/JHEP06(2013)072} {\bibfield  {journal} {\bibinfo  {journal} {JHEP}\
  }\textbf {\bibinfo {volume} {06}},\ \bibinfo {pages} {072} (\bibinfo {year}
  {2013})},\ \Eprint {http://arxiv.org/abs/1302.6216} {arXiv:1302.6216
  [hep-ph]} \BibitemShut {NoStop}%
\bibitem [{\citenamefont {Chen}\ \emph {et~al.}(2015)\citenamefont {Chen},
  \citenamefont {Gehrmann}, \citenamefont {Glover},\ and\ \citenamefont
  {Jaquier}}]{Chen:2014gva}%
  \BibitemOpen
  \bibfield  {author} {\bibinfo {author} {\bibfnamefont {X.}~\bibnamefont
  {Chen}}, \bibinfo {author} {\bibfnamefont {T.}~\bibnamefont {Gehrmann}},
  \bibinfo {author} {\bibfnamefont {E.~W.~N.}\ \bibnamefont {Glover}}, \ and\
  \bibinfo {author} {\bibfnamefont {M.}~\bibnamefont {Jaquier}},\ }\href
  {\doibase 10.1016/j.physletb.2014.11.021} {\bibfield  {journal} {\bibinfo
  {journal} {Phys. Lett.}\ }\textbf {\bibinfo {volume} {B740}},\ \bibinfo
  {pages} {147} (\bibinfo {year} {2015})},\ \Eprint
  {http://arxiv.org/abs/1408.5325} {arXiv:1408.5325 [hep-ph]} \BibitemShut
  {NoStop}%
\bibitem [{\citenamefont {Boughezal}\ \emph
  {et~al.}(2015{\natexlab{a}})\citenamefont {Boughezal}, \citenamefont {Caola},
  \citenamefont {Melnikov}, \citenamefont {Petriello},\ and\ \citenamefont
  {Schulze}}]{Boughezal:2015dra}%
  \BibitemOpen
  \bibfield  {author} {\bibinfo {author} {\bibfnamefont {R.}~\bibnamefont
  {Boughezal}}, \bibinfo {author} {\bibfnamefont {F.}~\bibnamefont {Caola}},
  \bibinfo {author} {\bibfnamefont {K.}~\bibnamefont {Melnikov}}, \bibinfo
  {author} {\bibfnamefont {F.}~\bibnamefont {Petriello}}, \ and\ \bibinfo
  {author} {\bibfnamefont {M.}~\bibnamefont {Schulze}},\ }\href {\doibase
  10.1103/PhysRevLett.115.082003} {\bibfield  {journal} {\bibinfo  {journal}
  {Phys. Rev. Lett.}\ }\textbf {\bibinfo {volume} {115}},\ \bibinfo {pages}
  {082003} (\bibinfo {year} {2015}{\natexlab{a}})},\ \Eprint
  {http://arxiv.org/abs/1504.07922} {arXiv:1504.07922 [hep-ph]} \BibitemShut
  {NoStop}%
\bibitem [{\citenamefont {Boughezal}\ \emph
  {et~al.}(2015{\natexlab{b}})\citenamefont {Boughezal}, \citenamefont {Focke},
  \citenamefont {Giele}, \citenamefont {Liu},\ and\ \citenamefont
  {Petriello}}]{Boughezal:2015aha}%
  \BibitemOpen
  \bibfield  {author} {\bibinfo {author} {\bibfnamefont {R.}~\bibnamefont
  {Boughezal}}, \bibinfo {author} {\bibfnamefont {C.}~\bibnamefont {Focke}},
  \bibinfo {author} {\bibfnamefont {W.}~\bibnamefont {Giele}}, \bibinfo
  {author} {\bibfnamefont {X.}~\bibnamefont {Liu}}, \ and\ \bibinfo {author}
  {\bibfnamefont {F.}~\bibnamefont {Petriello}},\ }\href {\doibase
  10.1016/j.physletb.2015.06.055} {\bibfield  {journal} {\bibinfo  {journal}
  {Phys. Lett.}\ }\textbf {\bibinfo {volume} {B748}},\ \bibinfo {pages} {5}
  (\bibinfo {year} {2015}{\natexlab{b}})},\ \Eprint
  {http://arxiv.org/abs/1505.03893} {arXiv:1505.03893 [hep-ph]} \BibitemShut
  {NoStop}%
\bibitem [{\citenamefont {Anastasiou}\ \emph
  {et~al.}(2016{\natexlab{a}})\citenamefont {Anastasiou}, \citenamefont {Duhr},
  \citenamefont {Dulat}, \citenamefont {Furlan}, \citenamefont {Gehrmann},
  \citenamefont {Herzog}, \citenamefont {Lazopoulos},\ and\ \citenamefont
  {Mistlberger}}]{Anastasiou:2016cez}%
  \BibitemOpen
  \bibfield  {author} {\bibinfo {author} {\bibfnamefont {C.}~\bibnamefont
  {Anastasiou}}, \bibinfo {author} {\bibfnamefont {C.}~\bibnamefont {Duhr}},
  \bibinfo {author} {\bibfnamefont {F.}~\bibnamefont {Dulat}}, \bibinfo
  {author} {\bibfnamefont {E.}~\bibnamefont {Furlan}}, \bibinfo {author}
  {\bibfnamefont {T.}~\bibnamefont {Gehrmann}}, \bibinfo {author}
  {\bibfnamefont {F.}~\bibnamefont {Herzog}}, \bibinfo {author} {\bibfnamefont
  {A.}~\bibnamefont {Lazopoulos}}, \ and\ \bibinfo {author} {\bibfnamefont
  {B.}~\bibnamefont {Mistlberger}},\ }\href {\doibase 10.1007/JHEP05(2016)058}
  {\bibfield  {journal} {\bibinfo  {journal} {JHEP}\ }\textbf {\bibinfo
  {volume} {05}},\ \bibinfo {pages} {058} (\bibinfo {year}
  {2016}{\natexlab{a}})},\ \Eprint {http://arxiv.org/abs/1602.00695}
  {arXiv:1602.00695 [hep-ph]} \BibitemShut {NoStop}%
\bibitem [{\citenamefont {Harlander}\ \emph {et~al.}(2017)\citenamefont
  {Harlander}, \citenamefont {Liebler},\ and\ \citenamefont
  {Mantler}}]{Harlander:2016hcx}%
  \BibitemOpen
  \bibfield  {author} {\bibinfo {author} {\bibfnamefont {R.~V.}\ \bibnamefont
  {Harlander}}, \bibinfo {author} {\bibfnamefont {S.}~\bibnamefont {Liebler}},
  \ and\ \bibinfo {author} {\bibfnamefont {H.}~\bibnamefont {Mantler}},\ }\href
  {\doibase 10.1016/j.cpc.2016.10.015} {\bibfield  {journal} {\bibinfo
  {journal} {Comput. Phys. Commun.}\ }\textbf {\bibinfo {volume} {212}},\
  \bibinfo {pages} {239} (\bibinfo {year} {2017})},\ \Eprint
  {http://arxiv.org/abs/1605.03190} {arXiv:1605.03190 [hep-ph]} \BibitemShut
  {NoStop}%
\bibitem [{\citenamefont {Anastasiou}\ \emph
  {et~al.}(2016{\natexlab{b}})\citenamefont {Anastasiou}, \citenamefont {Duhr},
  \citenamefont {Dulat}, \citenamefont {Furlan}, \citenamefont {Gehrmann},
  \citenamefont {Herzog}, \citenamefont {Lazopoulos},\ and\ \citenamefont
  {Mistlberger}}]{Anastasiou:2016hlm}%
  \BibitemOpen
  \bibfield  {author} {\bibinfo {author} {\bibfnamefont {C.}~\bibnamefont
  {Anastasiou}}, \bibinfo {author} {\bibfnamefont {C.}~\bibnamefont {Duhr}},
  \bibinfo {author} {\bibfnamefont {F.}~\bibnamefont {Dulat}}, \bibinfo
  {author} {\bibfnamefont {E.}~\bibnamefont {Furlan}}, \bibinfo {author}
  {\bibfnamefont {T.}~\bibnamefont {Gehrmann}}, \bibinfo {author}
  {\bibfnamefont {F.}~\bibnamefont {Herzog}}, \bibinfo {author} {\bibfnamefont
  {A.}~\bibnamefont {Lazopoulos}}, \ and\ \bibinfo {author} {\bibfnamefont
  {B.}~\bibnamefont {Mistlberger}},\ }\href {\doibase 10.1007/JHEP09(2016)037}
  {\bibfield  {journal} {\bibinfo  {journal} {JHEP}\ }\textbf {\bibinfo
  {volume} {09}},\ \bibinfo {pages} {037} (\bibinfo {year}
  {2016}{\natexlab{b}})},\ \Eprint {http://arxiv.org/abs/1605.05761}
  {arXiv:1605.05761 [hep-ph]} \BibitemShut {NoStop}%
\bibitem [{\citenamefont {Chen}\ \emph {et~al.}(2016)\citenamefont {Chen},
  \citenamefont {Cruz-Martinez}, \citenamefont {Gehrmann}, \citenamefont
  {Glover},\ and\ \citenamefont {Jaquier}}]{Chen:2016zka}%
  \BibitemOpen
  \bibfield  {author} {\bibinfo {author} {\bibfnamefont {X.}~\bibnamefont
  {Chen}}, \bibinfo {author} {\bibfnamefont {J.}~\bibnamefont {Cruz-Martinez}},
  \bibinfo {author} {\bibfnamefont {T.}~\bibnamefont {Gehrmann}}, \bibinfo
  {author} {\bibfnamefont {E.~W.~N.}\ \bibnamefont {Glover}}, \ and\ \bibinfo
  {author} {\bibfnamefont {M.}~\bibnamefont {Jaquier}},\ }\href {\doibase
  10.1007/JHEP10(2016)066} {\bibfield  {journal} {\bibinfo  {journal} {JHEP}\
  }\textbf {\bibinfo {volume} {10}},\ \bibinfo {pages} {066} (\bibinfo {year}
  {2016})},\ \Eprint {http://arxiv.org/abs/1607.08817} {arXiv:1607.08817
  [hep-ph]} \BibitemShut {NoStop}%
\bibitem [{\citenamefont {Lindert}\ \emph {et~al.}(2018)\citenamefont
  {Lindert}, \citenamefont {Kudashkin}, \citenamefont {Melnikov},\ and\
  \citenamefont {Wever}}]{Lindert:2018iug}%
  \BibitemOpen
  \bibfield  {author} {\bibinfo {author} {\bibfnamefont {J.~M.}\ \bibnamefont
  {Lindert}}, \bibinfo {author} {\bibfnamefont {K.}~\bibnamefont {Kudashkin}},
  \bibinfo {author} {\bibfnamefont {K.}~\bibnamefont {Melnikov}}, \ and\
  \bibinfo {author} {\bibfnamefont {C.}~\bibnamefont {Wever}},\ }\href@noop {}
  {\  (\bibinfo {year} {2018})},\ \Eprint {http://arxiv.org/abs/1801.08226}
  {arXiv:1801.08226 [hep-ph]} \BibitemShut {NoStop}%
\bibitem [{\citenamefont {Jones}\ \emph {et~al.}(2018)\citenamefont {Jones},
  \citenamefont {Kerner},\ and\ \citenamefont {Luisoni}}]{Jones:2018hbb}%
  \BibitemOpen
  \bibfield  {author} {\bibinfo {author} {\bibfnamefont {S.~P.}\ \bibnamefont
  {Jones}}, \bibinfo {author} {\bibfnamefont {M.}~\bibnamefont {Kerner}}, \
  and\ \bibinfo {author} {\bibfnamefont {G.}~\bibnamefont {Luisoni}},\
  }\href@noop {} {\  (\bibinfo {year} {2018})},\ \Eprint
  {http://arxiv.org/abs/1802.00349} {arXiv:1802.00349 [hep-ph]} \BibitemShut
  {NoStop}%
\bibitem [{\citenamefont {Neumann}(2018)}]{Neumann:2018bsx}%
  \BibitemOpen
  \bibfield  {author} {\bibinfo {author} {\bibfnamefont {T.}~\bibnamefont
  {Neumann}},\ }\href@noop {} {\  (\bibinfo {year} {2018})},\ \Eprint
  {http://arxiv.org/abs/1802.02981} {arXiv:1802.02981 [hep-ph]} \BibitemShut
  {NoStop}%
\bibitem [{\citenamefont {Kotikov}\ and\ \citenamefont
  {Lipatov}(2003)}]{Kotikov:2002ab}%
  \BibitemOpen
  \bibfield  {author} {\bibinfo {author} {\bibfnamefont {A.~V.}\ \bibnamefont
  {Kotikov}}\ and\ \bibinfo {author} {\bibfnamefont {L.~N.}\ \bibnamefont
  {Lipatov}},\ }\href {\doibase 10.1016/S0550-3213(03)00264-5,
  10.1016/j.nuclphysb.2004.02.032} {\bibfield  {journal} {\bibinfo  {journal}
  {Nucl. Phys.}\ }\textbf {\bibinfo {volume} {B661}},\ \bibinfo {pages} {19}
  (\bibinfo {year} {2003})},\ \bibinfo {note} {[Erratum: Nucl.
  Phys.B685,405(2004)]},\ \Eprint {http://arxiv.org/abs/hep-ph/0208220}
  {arXiv:hep-ph/0208220 [hep-ph]} \BibitemShut {NoStop}%
\bibitem [{\citenamefont {Kotikov}\ \emph {et~al.}(2004)\citenamefont
  {Kotikov}, \citenamefont {Lipatov}, \citenamefont {Onishchenko},\ and\
  \citenamefont {Velizhanin}}]{Kotikov:2004er}%
  \BibitemOpen
  \bibfield  {author} {\bibinfo {author} {\bibfnamefont {A.}~\bibnamefont
  {Kotikov}}, \bibinfo {author} {\bibfnamefont {L.}~\bibnamefont {Lipatov}},
  \bibinfo {author} {\bibfnamefont {A.}~\bibnamefont {Onishchenko}}, \ and\
  \bibinfo {author} {\bibfnamefont {V.}~\bibnamefont {Velizhanin}},\ }\href
  {\doibase 10.1016/j.physletb.2004.05.078, 10.1016/j.physletb.2004.05.078}
  {\bibfield  {journal} {\bibinfo  {journal} {Phys.Lett.}\ }\textbf {\bibinfo
  {volume} {B595}},\ \bibinfo {pages} {521} (\bibinfo {year} {2004})},\ \Eprint
  {http://arxiv.org/abs/hep-th/0404092} {arXiv:hep-th/0404092 [hep-th]}
  \BibitemShut {NoStop}%
\bibitem [{\citenamefont {Moch}\ \emph {et~al.}(2004)\citenamefont {Moch},
  \citenamefont {Vermaseren},\ and\ \citenamefont {Vogt}}]{Moch:2004pa}%
  \BibitemOpen
  \bibfield  {author} {\bibinfo {author} {\bibfnamefont {S.}~\bibnamefont
  {Moch}}, \bibinfo {author} {\bibfnamefont {J.~A.~M.}\ \bibnamefont
  {Vermaseren}}, \ and\ \bibinfo {author} {\bibfnamefont {A.}~\bibnamefont
  {Vogt}},\ }\href {\doibase 10.1016/j.nuclphysb.2004.03.030} {\bibfield
  {journal} {\bibinfo  {journal} {Nucl. Phys.}\ }\textbf {\bibinfo {volume}
  {B688}},\ \bibinfo {pages} {101} (\bibinfo {year} {2004})},\ \Eprint
  {http://arxiv.org/abs/hep-ph/0403192} {arXiv:hep-ph/0403192 [hep-ph]}
  \BibitemShut {NoStop}%
\bibitem [{\citenamefont {Brandhuber}\ \emph {et~al.}(2012)\citenamefont
  {Brandhuber}, \citenamefont {Travaglini},\ and\ \citenamefont
  {Yang}}]{Brandhuber:2012vm}%
  \BibitemOpen
  \bibfield  {author} {\bibinfo {author} {\bibfnamefont {A.}~\bibnamefont
  {Brandhuber}}, \bibinfo {author} {\bibfnamefont {G.}~\bibnamefont
  {Travaglini}}, \ and\ \bibinfo {author} {\bibfnamefont {G.}~\bibnamefont
  {Yang}},\ }\href@noop {} {\bibfield  {journal} {\bibinfo  {journal} {JHEP}\
  }\textbf {\bibinfo {volume} {1205}},\ \bibinfo {pages} {082} (\bibinfo {year}
  {2012})},\ \Eprint {http://arxiv.org/abs/1201.4170} {arXiv:1201.4170
  [hep-th]} \BibitemShut {NoStop}%
\bibitem [{\citenamefont {Gehrmann}\ \emph {et~al.}(2012)\citenamefont
  {Gehrmann}, \citenamefont {Jaquier}, \citenamefont {Glover},\ and\
  \citenamefont {Koukoutsakis}}]{Gehrmann:2011aa}%
  \BibitemOpen
  \bibfield  {author} {\bibinfo {author} {\bibfnamefont {T.}~\bibnamefont
  {Gehrmann}}, \bibinfo {author} {\bibfnamefont {M.}~\bibnamefont {Jaquier}},
  \bibinfo {author} {\bibfnamefont {E.}~\bibnamefont {Glover}}, \ and\ \bibinfo
  {author} {\bibfnamefont {A.}~\bibnamefont {Koukoutsakis}},\ }\href {\doibase
  10.1007/JHEP02(2012)056} {\bibfield  {journal} {\bibinfo  {journal} {JHEP}\
  }\textbf {\bibinfo {volume} {1202}},\ \bibinfo {pages} {056} (\bibinfo {year}
  {2012})},\ \Eprint {http://arxiv.org/abs/1112.3554} {arXiv:1112.3554
  [hep-ph]} \BibitemShut {NoStop}%
\bibitem [{\citenamefont {Jin}\ and\ \citenamefont {Yang}(2018)}]{Jin:2018fak}%
  \BibitemOpen
  \bibfield  {author} {\bibinfo {author} {\bibfnamefont {Q.}~\bibnamefont
  {Jin}}\ and\ \bibinfo {author} {\bibfnamefont {G.}~\bibnamefont {Yang}},\
  }\href {\doibase 10.1103/PhysRevLett.121.101603} {\bibfield  {journal}
  {\bibinfo  {journal} {Phys. Rev. Lett.}\ }\textbf {\bibinfo {volume} {121}},\
  \bibinfo {pages} {101603} (\bibinfo {year} {2018})},\ \Eprint
  {http://arxiv.org/abs/1804.04653} {arXiv:1804.04653 [hep-th]} \BibitemShut
  {NoStop}%
\bibitem [{\citenamefont {Brandhuber}\ \emph {et~al.}(2017)\citenamefont
  {Brandhuber}, \citenamefont {Kostacinska}, \citenamefont {Penante},\ and\
  \citenamefont {Travaglini}}]{Brandhuber:2017bkg}%
  \BibitemOpen
  \bibfield  {author} {\bibinfo {author} {\bibfnamefont {A.}~\bibnamefont
  {Brandhuber}}, \bibinfo {author} {\bibfnamefont {M.}~\bibnamefont
  {Kostacinska}}, \bibinfo {author} {\bibfnamefont {B.}~\bibnamefont
  {Penante}}, \ and\ \bibinfo {author} {\bibfnamefont {G.}~\bibnamefont
  {Travaglini}},\ }\href {\doibase 10.1103/PhysRevLett.119.161601} {\bibfield
  {journal} {\bibinfo  {journal} {Phys. Rev. Lett.}\ }\textbf {\bibinfo
  {volume} {119}},\ \bibinfo {pages} {161601} (\bibinfo {year} {2017})},\
  \Eprint {http://arxiv.org/abs/1707.09897} {arXiv:1707.09897 [hep-th]}
  \BibitemShut {NoStop}%
\bibitem [{\citenamefont {Brandhuber}\ \emph
  {et~al.}(2018{\natexlab{a}})\citenamefont {Brandhuber}, \citenamefont
  {Kostacinska}, \citenamefont {Penante},\ and\ \citenamefont
  {Travaglini}}]{Brandhuber:2018xzk}%
  \BibitemOpen
  \bibfield  {author} {\bibinfo {author} {\bibfnamefont {A.}~\bibnamefont
  {Brandhuber}}, \bibinfo {author} {\bibfnamefont {M.}~\bibnamefont
  {Kostacinska}}, \bibinfo {author} {\bibfnamefont {B.}~\bibnamefont
  {Penante}}, \ and\ \bibinfo {author} {\bibfnamefont {G.}~\bibnamefont
  {Travaglini}},\ }\href {\doibase 10.1007/JHEP12(2018)076} {\bibfield
  {journal} {\bibinfo  {journal} {JHEP}\ }\textbf {\bibinfo {volume} {12}},\
  \bibinfo {pages} {076} (\bibinfo {year} {2018}{\natexlab{a}})},\ \Eprint
  {http://arxiv.org/abs/1804.05703} {arXiv:1804.05703 [hep-th]} \BibitemShut
  {NoStop}%
\bibitem [{\citenamefont {Brandhuber}\ \emph
  {et~al.}(2018{\natexlab{b}})\citenamefont {Brandhuber}, \citenamefont
  {Kostacinska}, \citenamefont {Penante},\ and\ \citenamefont
  {Travaglini}}]{Brandhuber:2018kqb}%
  \BibitemOpen
  \bibfield  {author} {\bibinfo {author} {\bibfnamefont {A.}~\bibnamefont
  {Brandhuber}}, \bibinfo {author} {\bibfnamefont {M.}~\bibnamefont
  {Kostacinska}}, \bibinfo {author} {\bibfnamefont {B.}~\bibnamefont
  {Penante}}, \ and\ \bibinfo {author} {\bibfnamefont {G.}~\bibnamefont
  {Travaglini}},\ }\href {\doibase 10.1007/JHEP12(2018)077} {\bibfield
  {journal} {\bibinfo  {journal} {JHEP}\ }\textbf {\bibinfo {volume} {12}},\
  \bibinfo {pages} {077} (\bibinfo {year} {2018}{\natexlab{b}})},\ \Eprint
  {http://arxiv.org/abs/1804.05828} {arXiv:1804.05828 [hep-th]} \BibitemShut
  {NoStop}%
\bibitem [{\citenamefont {Li}\ \emph {et~al.}(2015)\citenamefont {Li},
  \citenamefont {von Manteuffel}, \citenamefont {Schabinger},\ and\
  \citenamefont {Zhu}}]{Li:2014afw}%
  \BibitemOpen
  \bibfield  {author} {\bibinfo {author} {\bibfnamefont {Y.}~\bibnamefont
  {Li}}, \bibinfo {author} {\bibfnamefont {A.}~\bibnamefont {von Manteuffel}},
  \bibinfo {author} {\bibfnamefont {R.~M.}\ \bibnamefont {Schabinger}}, \ and\
  \bibinfo {author} {\bibfnamefont {H.~X.}\ \bibnamefont {Zhu}},\ }\href
  {\doibase 10.1103/PhysRevD.91.036008} {\bibfield  {journal} {\bibinfo
  {journal} {Phys. Rev.}\ }\textbf {\bibinfo {volume} {D91}},\ \bibinfo {pages}
  {036008} (\bibinfo {year} {2015})},\ \Eprint {http://arxiv.org/abs/1412.2771}
  {arXiv:1412.2771 [hep-ph]} \BibitemShut {NoStop}%
\bibitem [{\citenamefont {Li}\ and\ \citenamefont {Zhu}(2017)}]{Li:2016ctv}%
  \BibitemOpen
  \bibfield  {author} {\bibinfo {author} {\bibfnamefont {Y.}~\bibnamefont
  {Li}}\ and\ \bibinfo {author} {\bibfnamefont {H.~X.}\ \bibnamefont {Zhu}},\
  }\href {\doibase 10.1103/PhysRevLett.118.022004} {\bibfield  {journal}
  {\bibinfo  {journal} {Phys. Rev. Lett.}\ }\textbf {\bibinfo {volume} {118}},\
  \bibinfo {pages} {022004} (\bibinfo {year} {2017})},\ \Eprint
  {http://arxiv.org/abs/1604.01404} {arXiv:1604.01404 [hep-ph]} \BibitemShut
  {NoStop}%
\bibitem [{\citenamefont {Dixon}(2018)}]{Dixon:2017nat}%
  \BibitemOpen
  \bibfield  {author} {\bibinfo {author} {\bibfnamefont {L.~J.}\ \bibnamefont
  {Dixon}},\ }\href {\doibase 10.1007/JHEP01(2018)075} {\bibfield  {journal}
  {\bibinfo  {journal} {JHEP}\ }\textbf {\bibinfo {volume} {01}},\ \bibinfo
  {pages} {075} (\bibinfo {year} {2018})},\ \Eprint
  {http://arxiv.org/abs/1712.07274} {arXiv:1712.07274 [hep-th]} \BibitemShut
  {NoStop}%
\bibitem [{\citenamefont {Brandhuber}\ \emph {et~al.}(2014)\citenamefont
  {Brandhuber}, \citenamefont {Penante}, \citenamefont {Travaglini},\ and\
  \citenamefont {Wen}}]{Brandhuber:2014ica}%
  \BibitemOpen
  \bibfield  {author} {\bibinfo {author} {\bibfnamefont {A.}~\bibnamefont
  {Brandhuber}}, \bibinfo {author} {\bibfnamefont {B.}~\bibnamefont {Penante}},
  \bibinfo {author} {\bibfnamefont {G.}~\bibnamefont {Travaglini}}, \ and\
  \bibinfo {author} {\bibfnamefont {C.}~\bibnamefont {Wen}},\ }\href {\doibase
  10.1007/JHEP08(2014)100} {\bibfield  {journal} {\bibinfo  {journal} {JHEP}\
  }\textbf {\bibinfo {volume} {08}},\ \bibinfo {pages} {100} (\bibinfo {year}
  {2014})},\ \Eprint {http://arxiv.org/abs/1406.1443} {arXiv:1406.1443
  [hep-th]} \BibitemShut {NoStop}%
\bibitem [{\citenamefont {Loebbert}\ \emph {et~al.}(2015)\citenamefont
  {Loebbert}, \citenamefont {Nandan}, \citenamefont {Sieg}, \citenamefont
  {Wilhelm},\ and\ \citenamefont {Yang}}]{Loebbert:2015ova}%
  \BibitemOpen
  \bibfield  {author} {\bibinfo {author} {\bibfnamefont {F.}~\bibnamefont
  {Loebbert}}, \bibinfo {author} {\bibfnamefont {D.}~\bibnamefont {Nandan}},
  \bibinfo {author} {\bibfnamefont {C.}~\bibnamefont {Sieg}}, \bibinfo {author}
  {\bibfnamefont {M.}~\bibnamefont {Wilhelm}}, \ and\ \bibinfo {author}
  {\bibfnamefont {G.}~\bibnamefont {Yang}},\ }\href {\doibase
  10.1007/JHEP10(2015)012} {\bibfield  {journal} {\bibinfo  {journal} {JHEP}\
  }\textbf {\bibinfo {volume} {10}},\ \bibinfo {pages} {012} (\bibinfo {year}
  {2015})},\ \Eprint {http://arxiv.org/abs/1504.06323} {arXiv:1504.06323
  [hep-th]} \BibitemShut {NoStop}%
\bibitem [{\citenamefont {Brandhuber}\ \emph {et~al.}(2016)\citenamefont
  {Brandhuber}, \citenamefont {Kostacinska}, \citenamefont {Penante},
  \citenamefont {Travaglini},\ and\ \citenamefont
  {Young}}]{Brandhuber:2016fni}%
  \BibitemOpen
  \bibfield  {author} {\bibinfo {author} {\bibfnamefont {A.}~\bibnamefont
  {Brandhuber}}, \bibinfo {author} {\bibfnamefont {M.}~\bibnamefont
  {Kostacinska}}, \bibinfo {author} {\bibfnamefont {B.}~\bibnamefont
  {Penante}}, \bibinfo {author} {\bibfnamefont {G.}~\bibnamefont {Travaglini}},
  \ and\ \bibinfo {author} {\bibfnamefont {D.}~\bibnamefont {Young}},\ }\href
  {\doibase 10.1007/JHEP08(2016)134} {\bibfield  {journal} {\bibinfo  {journal}
  {JHEP}\ }\textbf {\bibinfo {volume} {08}},\ \bibinfo {pages} {134} (\bibinfo
  {year} {2016})},\ \Eprint {http://arxiv.org/abs/1606.08682} {arXiv:1606.08682
  [hep-th]} \BibitemShut {NoStop}%
\bibitem [{\citenamefont {Loebbert}\ \emph {et~al.}(2016)\citenamefont
  {Loebbert}, \citenamefont {Sieg}, \citenamefont {Wilhelm},\ and\
  \citenamefont {Yang}}]{Loebbert:2016xkw}%
  \BibitemOpen
  \bibfield  {author} {\bibinfo {author} {\bibfnamefont {F.}~\bibnamefont
  {Loebbert}}, \bibinfo {author} {\bibfnamefont {C.}~\bibnamefont {Sieg}},
  \bibinfo {author} {\bibfnamefont {M.}~\bibnamefont {Wilhelm}}, \ and\
  \bibinfo {author} {\bibfnamefont {G.}~\bibnamefont {Yang}},\ }\href {\doibase
  10.1007/JHEP12(2016)090} {\bibfield  {journal} {\bibinfo  {journal} {JHEP}\
  }\textbf {\bibinfo {volume} {12}},\ \bibinfo {pages} {090} (\bibinfo {year}
  {2016})},\ \Eprint {http://arxiv.org/abs/1610.06567} {arXiv:1610.06567
  [hep-th]} \BibitemShut {NoStop}%
\bibitem [{\citenamefont {Buchmuller}\ and\ \citenamefont
  {Wyler}(1986)}]{Buchmuller:1985jz}%
  \BibitemOpen
  \bibfield  {author} {\bibinfo {author} {\bibfnamefont {W.}~\bibnamefont
  {Buchmuller}}\ and\ \bibinfo {author} {\bibfnamefont {D.}~\bibnamefont
  {Wyler}},\ }\href {\doibase 10.1016/0550-3213(86)90262-2} {\bibfield
  {journal} {\bibinfo  {journal} {Nucl. Phys.}\ }\textbf {\bibinfo {volume}
  {B268}},\ \bibinfo {pages} {621} (\bibinfo {year} {1986})}\BibitemShut
  {NoStop}%
\bibitem [{\citenamefont {Gracey}(2002)}]{Gracey:2002he}%
  \BibitemOpen
  \bibfield  {author} {\bibinfo {author} {\bibfnamefont {J.~A.}\ \bibnamefont
  {Gracey}},\ }\href {\doibase 10.1016/S0550-3213(02)00334-6,
  10.1016/j.nuclphysb.2004.06.053} {\bibfield  {journal} {\bibinfo  {journal}
  {Nucl. Phys.}\ }\textbf {\bibinfo {volume} {B634}},\ \bibinfo {pages} {192}
  (\bibinfo {year} {2002})},\ \bibinfo {note} {[Erratum: Nucl.
  Phys.B696,295(2004)]},\ \Eprint {http://arxiv.org/abs/hep-ph/0204266}
  {arXiv:hep-ph/0204266 [hep-ph]} \BibitemShut {NoStop}%
\bibitem [{\citenamefont {Neill}(2009)}]{Neill:2009tn}%
  \BibitemOpen
  \bibfield  {author} {\bibinfo {author} {\bibfnamefont {D.}~\bibnamefont
  {Neill}},\ }\href@noop {} {\  (\bibinfo {year} {2009})},\ \Eprint
  {http://arxiv.org/abs/0908.1573} {arXiv:0908.1573 [hep-ph]} \BibitemShut
  {NoStop}%
\bibitem [{\citenamefont {Harlander}\ and\ \citenamefont
  {Neumann}(2013)}]{Harlander:2013oja}%
  \BibitemOpen
  \bibfield  {author} {\bibinfo {author} {\bibfnamefont {R.~V.}\ \bibnamefont
  {Harlander}}\ and\ \bibinfo {author} {\bibfnamefont {T.}~\bibnamefont
  {Neumann}},\ }\href {\doibase 10.1103/PhysRevD.88.074015} {\bibfield
  {journal} {\bibinfo  {journal} {Phys. Rev.}\ }\textbf {\bibinfo {volume}
  {D88}},\ \bibinfo {pages} {074015} (\bibinfo {year} {2013})},\ \Eprint
  {http://arxiv.org/abs/1308.2225} {arXiv:1308.2225 [hep-ph]} \BibitemShut
  {NoStop}%
\bibitem [{\citenamefont {Dawson}\ \emph {et~al.}(2014)\citenamefont {Dawson},
  \citenamefont {Lewis},\ and\ \citenamefont {Zeng}}]{Dawson:2014ora}%
  \BibitemOpen
  \bibfield  {author} {\bibinfo {author} {\bibfnamefont {S.}~\bibnamefont
  {Dawson}}, \bibinfo {author} {\bibfnamefont {I.~M.}\ \bibnamefont {Lewis}}, \
  and\ \bibinfo {author} {\bibfnamefont {M.}~\bibnamefont {Zeng}},\ }\href
  {\doibase 10.1103/PhysRevD.90.093007} {\bibfield  {journal} {\bibinfo
  {journal} {Phys. Rev.}\ }\textbf {\bibinfo {volume} {D90}},\ \bibinfo {pages}
  {093007} (\bibinfo {year} {2014})},\ \Eprint {http://arxiv.org/abs/1409.6299}
  {arXiv:1409.6299 [hep-ph]} \BibitemShut {NoStop}%
\bibitem [{\citenamefont {Bern}\ \emph {et~al.}(1994)\citenamefont {Bern},
  \citenamefont {Dixon}, \citenamefont {Dunbar},\ and\ \citenamefont
  {Kosower}}]{Bern:1994zx}%
  \BibitemOpen
  \bibfield  {author} {\bibinfo {author} {\bibfnamefont {Z.}~\bibnamefont
  {Bern}}, \bibinfo {author} {\bibfnamefont {L.~J.}\ \bibnamefont {Dixon}},
  \bibinfo {author} {\bibfnamefont {D.~C.}\ \bibnamefont {Dunbar}}, \ and\
  \bibinfo {author} {\bibfnamefont {D.~A.}\ \bibnamefont {Kosower}},\ }\href
  {\doibase 10.1016/0550-3213(94)90179-1} {\bibfield  {journal} {\bibinfo
  {journal} {Nucl.Phys.}\ }\textbf {\bibinfo {volume} {B425}},\ \bibinfo
  {pages} {217} (\bibinfo {year} {1994})},\ \Eprint
  {http://arxiv.org/abs/hep-ph/9403226} {arXiv:hep-ph/9403226 [hep-ph]}
  \BibitemShut {NoStop}%
\bibitem [{\citenamefont {Bern}\ \emph {et~al.}(1995)\citenamefont {Bern},
  \citenamefont {Dixon}, \citenamefont {Dunbar},\ and\ \citenamefont
  {Kosower}}]{Bern:1994cg}%
  \BibitemOpen
  \bibfield  {author} {\bibinfo {author} {\bibfnamefont {Z.}~\bibnamefont
  {Bern}}, \bibinfo {author} {\bibfnamefont {L.~J.}\ \bibnamefont {Dixon}},
  \bibinfo {author} {\bibfnamefont {D.~C.}\ \bibnamefont {Dunbar}}, \ and\
  \bibinfo {author} {\bibfnamefont {D.~A.}\ \bibnamefont {Kosower}},\ }\href
  {\doibase 10.1016/0550-3213(94)00488-Z} {\bibfield  {journal} {\bibinfo
  {journal} {Nucl. Phys.}\ }\textbf {\bibinfo {volume} {B435}},\ \bibinfo
  {pages} {59} (\bibinfo {year} {1995})},\ \Eprint
  {http://arxiv.org/abs/hep-ph/9409265} {arXiv:hep-ph/9409265 [hep-ph]}
  \BibitemShut {NoStop}%
\bibitem [{\citenamefont {Britto}\ \emph {et~al.}(2005)\citenamefont {Britto},
  \citenamefont {Cachazo},\ and\ \citenamefont {Feng}}]{Britto:2004nc}%
  \BibitemOpen
  \bibfield  {author} {\bibinfo {author} {\bibfnamefont {R.}~\bibnamefont
  {Britto}}, \bibinfo {author} {\bibfnamefont {F.}~\bibnamefont {Cachazo}}, \
  and\ \bibinfo {author} {\bibfnamefont {B.}~\bibnamefont {Feng}},\ }\href
  {\doibase 10.1016/j.nuclphysb.2005.07.014} {\bibfield  {journal} {\bibinfo
  {journal} {Nucl.Phys.}\ }\textbf {\bibinfo {volume} {B725}},\ \bibinfo
  {pages} {275} (\bibinfo {year} {2005})},\ \Eprint
  {http://arxiv.org/abs/hep-th/0412103} {arXiv:hep-th/0412103 [hep-th]}
  \BibitemShut {NoStop}%
\bibitem [{\citenamefont {Chetyrkin}\ and\ \citenamefont
  {Tkachov}(1981)}]{Chetyrkin:1981qh}%
  \BibitemOpen
  \bibfield  {author} {\bibinfo {author} {\bibfnamefont {K.}~\bibnamefont
  {Chetyrkin}}\ and\ \bibinfo {author} {\bibfnamefont {F.}~\bibnamefont
  {Tkachov}},\ }\href {\doibase 10.1016/0550-3213(81)90199-1} {\bibfield
  {journal} {\bibinfo  {journal} {Nucl.Phys.}\ }\textbf {\bibinfo {volume}
  {B192}},\ \bibinfo {pages} {159} (\bibinfo {year} {1981})}\BibitemShut
  {NoStop}%
\bibitem [{\citenamefont {Tkachov}(1981)}]{Tkachov:1981wb}%
  \BibitemOpen
  \bibfield  {author} {\bibinfo {author} {\bibfnamefont {F.}~\bibnamefont
  {Tkachov}},\ }\href {\doibase 10.1016/0370-2693(81)90288-4} {\bibfield
  {journal} {\bibinfo  {journal} {Phys.Lett.}\ }\textbf {\bibinfo {volume}
  {B100}},\ \bibinfo {pages} {65} (\bibinfo {year} {1981})}\BibitemShut
  {NoStop}%
\bibitem [{\citenamefont {Hahn}(2001)}]{Hahn:2000kx}%
  \BibitemOpen
  \bibfield  {author} {\bibinfo {author} {\bibfnamefont {T.}~\bibnamefont
  {Hahn}},\ }\href {\doibase 10.1016/S0010-4655(01)00290-9} {\bibfield
  {journal} {\bibinfo  {journal} {Comput. Phys. Commun.}\ }\textbf {\bibinfo
  {volume} {140}},\ \bibinfo {pages} {418} (\bibinfo {year} {2001})},\ \Eprint
  {http://arxiv.org/abs/hep-ph/0012260} {arXiv:hep-ph/0012260 [hep-ph]}
  \BibitemShut {NoStop}%
\bibitem [{\citenamefont {Boels}\ and\ \citenamefont
  {Luo}(2017)}]{Boels:2017gyc}%
  \BibitemOpen
  \bibfield  {author} {\bibinfo {author} {\bibfnamefont {R.~H.}\ \bibnamefont
  {Boels}}\ and\ \bibinfo {author} {\bibfnamefont {H.}~\bibnamefont {Luo}},\
  }\href@noop {} {\  (\bibinfo {year} {2017})},\ \Eprint
  {http://arxiv.org/abs/1710.10208} {arXiv:1710.10208 [hep-th]} \BibitemShut
  {NoStop}%
\bibitem [{\citenamefont {Boels}\ \emph {et~al.}(2018)\citenamefont {Boels},
  \citenamefont {Jin},\ and\ \citenamefont {Luo}}]{Boels:2018nrr}%
  \BibitemOpen
  \bibfield  {author} {\bibinfo {author} {\bibfnamefont {R.~H.}\ \bibnamefont
  {Boels}}, \bibinfo {author} {\bibfnamefont {Q.}~\bibnamefont {Jin}}, \ and\
  \bibinfo {author} {\bibfnamefont {H.}~\bibnamefont {Luo}},\ }\href@noop {} {\
   (\bibinfo {year} {2018})},\ \Eprint {http://arxiv.org/abs/1802.06761}
  {arXiv:1802.06761 [hep-ph]} \BibitemShut {NoStop}%
\bibitem [{\citenamefont {Smirnov}(2015)}]{Smirnov:2014hma}%
  \BibitemOpen
  \bibfield  {author} {\bibinfo {author} {\bibfnamefont {A.~V.}\ \bibnamefont
  {Smirnov}},\ }\href {\doibase 10.1016/j.cpc.2014.11.024} {\bibfield
  {journal} {\bibinfo  {journal} {Comput. Phys. Commun.}\ }\textbf {\bibinfo
  {volume} {189}},\ \bibinfo {pages} {182} (\bibinfo {year} {2015})},\ \Eprint
  {http://arxiv.org/abs/1408.2372} {arXiv:1408.2372 [hep-ph]} \BibitemShut
  {NoStop}%
\bibitem [{\citenamefont {Lee}(2014)}]{Lee:2013mka}%
  \BibitemOpen
  \bibfield  {author} {\bibinfo {author} {\bibfnamefont {R.~N.}\ \bibnamefont
  {Lee}},\ }\bibfield  {booktitle} {\emph {\bibinfo {booktitle} {{Proceedings,
  15th International Workshop on Advanced Computing and Analysis Techniques in
  Physics Research (ACAT 2013): Beijing, China, May 16-21, 2013}}},\ }\href
  {\doibase 10.1088/1742-6596/523/1/012059} {\bibfield  {journal} {\bibinfo
  {journal} {J. Phys. Conf. Ser.}\ }\textbf {\bibinfo {volume} {523}},\
  \bibinfo {pages} {012059} (\bibinfo {year} {2014})},\ \Eprint
  {http://arxiv.org/abs/1310.1145} {arXiv:1310.1145 [hep-ph]} \BibitemShut
  {NoStop}%
\bibitem [{\citenamefont {von Manteuffel}\ and\ \citenamefont
  {Studerus}(2012)}]{vonManteuffel:2012np}%
  \BibitemOpen
  \bibfield  {author} {\bibinfo {author} {\bibfnamefont {A.}~\bibnamefont {von
  Manteuffel}}\ and\ \bibinfo {author} {\bibfnamefont {C.}~\bibnamefont
  {Studerus}},\ }\href@noop {} {\  (\bibinfo {year} {2012})},\ \Eprint
  {http://arxiv.org/abs/1201.4330} {arXiv:1201.4330 [hep-ph]} \BibitemShut
  {NoStop}%
\bibitem [{\citenamefont {Maierhoefer}\ \emph {et~al.}(2017)\citenamefont
  {Maierhoefer}, \citenamefont {Usovitsch},\ and\ \citenamefont
  {Uwer}}]{Maierhoefer:2017hyi}%
  \BibitemOpen
  \bibfield  {author} {\bibinfo {author} {\bibfnamefont {P.}~\bibnamefont
  {Maierhoefer}}, \bibinfo {author} {\bibfnamefont {J.}~\bibnamefont
  {Usovitsch}}, \ and\ \bibinfo {author} {\bibfnamefont {P.}~\bibnamefont
  {Uwer}},\ }\href {\doibase 10.1016/j.cpc.2018.04.012} {\  (\bibinfo {year}
  {2017}),\ 10.1016/j.cpc.2018.04.012},\ \Eprint
  {http://arxiv.org/abs/1705.05610} {arXiv:1705.05610 [hep-ph]} \BibitemShut
  {NoStop}%
\bibitem [{\citenamefont {Gehrmann}\ and\ \citenamefont
  {Remiddi}(2001{\natexlab{a}})}]{Gehrmann:2000zt}%
  \BibitemOpen
  \bibfield  {author} {\bibinfo {author} {\bibfnamefont {T.}~\bibnamefont
  {Gehrmann}}\ and\ \bibinfo {author} {\bibfnamefont {E.}~\bibnamefont
  {Remiddi}},\ }\href {\doibase 10.1016/S0550-3213(01)00057-8} {\bibfield
  {journal} {\bibinfo  {journal} {Nucl. Phys.}\ }\textbf {\bibinfo {volume}
  {B601}},\ \bibinfo {pages} {248} (\bibinfo {year} {2001}{\natexlab{a}})},\
  \Eprint {http://arxiv.org/abs/hep-ph/0008287} {arXiv:hep-ph/0008287 [hep-ph]}
  \BibitemShut {NoStop}%
\bibitem [{\citenamefont {Gehrmann}\ and\ \citenamefont
  {Remiddi}(2001{\natexlab{b}})}]{Gehrmann:2001ck}%
  \BibitemOpen
  \bibfield  {author} {\bibinfo {author} {\bibfnamefont {T.}~\bibnamefont
  {Gehrmann}}\ and\ \bibinfo {author} {\bibfnamefont {E.}~\bibnamefont
  {Remiddi}},\ }\href {\doibase 10.1016/S0550-3213(01)00074-8} {\bibfield
  {journal} {\bibinfo  {journal} {Nucl. Phys.}\ }\textbf {\bibinfo {volume}
  {B601}},\ \bibinfo {pages} {287} (\bibinfo {year} {2001}{\natexlab{b}})},\
  \Eprint {http://arxiv.org/abs/hep-ph/0101124} {arXiv:hep-ph/0101124 [hep-ph]}
  \BibitemShut {NoStop}%
\bibitem [{\citenamefont {Gehrmann}\ and\ \citenamefont
  {Remiddi}(2002)}]{Gehrmann:2001jv}%
  \BibitemOpen
  \bibfield  {author} {\bibinfo {author} {\bibfnamefont {T.}~\bibnamefont
  {Gehrmann}}\ and\ \bibinfo {author} {\bibfnamefont {E.}~\bibnamefont
  {Remiddi}},\ }\href {\doibase 10.1016/S0010-4655(02)00139-X} {\bibfield
  {journal} {\bibinfo  {journal} {Comput. Phys. Commun.}\ }\textbf {\bibinfo
  {volume} {144}},\ \bibinfo {pages} {200} (\bibinfo {year} {2002})},\ \Eprint
  {http://arxiv.org/abs/hep-ph/0111255} {arXiv:hep-ph/0111255 [hep-ph]}
  \BibitemShut {NoStop}%
\bibitem [{\citenamefont {Bardeen}\ \emph {et~al.}(1978)\citenamefont
  {Bardeen}, \citenamefont {Buras}, \citenamefont {Duke},\ and\ \citenamefont
  {Muta}}]{Bardeen:1978yd}%
  \BibitemOpen
  \bibfield  {author} {\bibinfo {author} {\bibfnamefont {W.~A.}\ \bibnamefont
  {Bardeen}}, \bibinfo {author} {\bibfnamefont {A.~J.}\ \bibnamefont {Buras}},
  \bibinfo {author} {\bibfnamefont {D.~W.}\ \bibnamefont {Duke}}, \ and\
  \bibinfo {author} {\bibfnamefont {T.}~\bibnamefont {Muta}},\ }\href {\doibase
  10.1103/PhysRevD.18.3998} {\bibfield  {journal} {\bibinfo  {journal} {Phys.
  Rev.}\ }\textbf {\bibinfo {volume} {D18}},\ \bibinfo {pages} {3998} (\bibinfo
  {year} {1978})}\BibitemShut {NoStop}%
\bibitem [{\citenamefont {Catani}(1998)}]{Catani:1998bh}%
  \BibitemOpen
  \bibfield  {author} {\bibinfo {author} {\bibfnamefont {S.}~\bibnamefont
  {Catani}},\ }\href {\doibase 10.1016/S0370-2693(98)00332-3} {\bibfield
  {journal} {\bibinfo  {journal} {Phys. Lett.}\ }\textbf {\bibinfo {volume}
  {B427}},\ \bibinfo {pages} {161} (\bibinfo {year} {1998})},\ \Eprint
  {http://arxiv.org/abs/hep-ph/9802439} {arXiv:hep-ph/9802439 [hep-ph]}
  \BibitemShut {NoStop}%
\bibitem [{\citenamefont {Jin}\ and\ \citenamefont {Yang}()}]{supplemental}%
  \BibitemOpen
  \bibfield  {author} {\bibinfo {author} {\bibfnamefont {Q.}~\bibnamefont
  {Jin}}\ and\ \bibinfo {author} {\bibfnamefont {G.}~\bibnamefont {Yang}},\
  }\href@noop {} {\bibinfo  {journal} {Supplemental Material}\ }\BibitemShut
  {NoStop}%
\bibitem [{\citenamefont {Banerjee}\ \emph
  {et~al.}(2017{\natexlab{a}})\citenamefont {Banerjee}, \citenamefont {Dhani},
  \citenamefont {Mahakhud}, \citenamefont {Ravindran},\ and\ \citenamefont
  {Seth}}]{Banerjee:2016kri}%
  \BibitemOpen
\bibfield  {journal} {  }\bibfield  {author} {\bibinfo {author} {\bibfnamefont
  {P.}~\bibnamefont {Banerjee}}, \bibinfo {author} {\bibfnamefont {P.~K.}\
  \bibnamefont {Dhani}}, \bibinfo {author} {\bibfnamefont {M.}~\bibnamefont
  {Mahakhud}}, \bibinfo {author} {\bibfnamefont {V.}~\bibnamefont {Ravindran}},
  \ and\ \bibinfo {author} {\bibfnamefont {S.}~\bibnamefont {Seth}},\ }\href
  {\doibase 10.1007/JHEP05(2017)085} {\bibfield  {journal} {\bibinfo  {journal}
  {JHEP}\ }\textbf {\bibinfo {volume} {05}},\ \bibinfo {pages} {085} (\bibinfo
  {year} {2017}{\natexlab{a}})},\ \Eprint {http://arxiv.org/abs/1612.00885}
  {arXiv:1612.00885 [hep-th]} \BibitemShut {NoStop}%
\bibitem [{\citenamefont {Bern}\ \emph {et~al.}(2005)\citenamefont {Bern},
  \citenamefont {Dixon},\ and\ \citenamefont {Smirnov}}]{Bern:2005iz}%
  \BibitemOpen
  \bibfield  {author} {\bibinfo {author} {\bibfnamefont {Z.}~\bibnamefont
  {Bern}}, \bibinfo {author} {\bibfnamefont {L.~J.}\ \bibnamefont {Dixon}}, \
  and\ \bibinfo {author} {\bibfnamefont {V.~A.}\ \bibnamefont {Smirnov}},\
  }\href {\doibase 10.1103/PhysRevD.72.085001} {\bibfield  {journal} {\bibinfo
  {journal} {Phys. Rev.}\ }\textbf {\bibinfo {volume} {D72}},\ \bibinfo {pages}
  {085001} (\bibinfo {year} {2005})},\ \Eprint
  {http://arxiv.org/abs/hep-th/0505205} {arXiv:hep-th/0505205 [hep-th]}
  \BibitemShut {NoStop}%
\bibitem [{\citenamefont {Duhr}(2012)}]{Duhr:2012fh}%
  \BibitemOpen
  \bibfield  {author} {\bibinfo {author} {\bibfnamefont {C.}~\bibnamefont
  {Duhr}},\ }\href {\doibase 10.1007/JHEP08(2012)043} {\bibfield  {journal}
  {\bibinfo  {journal} {JHEP}\ }\textbf {\bibinfo {volume} {08}},\ \bibinfo
  {pages} {043} (\bibinfo {year} {2012})},\ \Eprint
  {http://arxiv.org/abs/1203.0454} {arXiv:1203.0454 [hep-ph]} \BibitemShut
  {NoStop}%
\bibitem [{\citenamefont {Banerjee}\ \emph
  {et~al.}(2017{\natexlab{b}})\citenamefont {Banerjee}, \citenamefont {Dhani},\
  and\ \citenamefont {Ravindran}}]{Banerjee:2017faz}%
  \BibitemOpen
  \bibfield  {author} {\bibinfo {author} {\bibfnamefont {P.}~\bibnamefont
  {Banerjee}}, \bibinfo {author} {\bibfnamefont {P.~K.}\ \bibnamefont {Dhani}},
  \ and\ \bibinfo {author} {\bibfnamefont {V.}~\bibnamefont {Ravindran}},\
  }\href {\doibase 10.1007/JHEP10(2017)067} {\bibfield  {journal} {\bibinfo
  {journal} {JHEP}\ }\textbf {\bibinfo {volume} {10}},\ \bibinfo {pages} {067}
  (\bibinfo {year} {2017}{\natexlab{b}})},\ \Eprint
  {http://arxiv.org/abs/1708.02387} {arXiv:1708.02387 [hep-ph]} \BibitemShut
  {NoStop}%
\bibitem [{\citenamefont {Dixon}\ and\ \citenamefont
  {Sofianatos}(2009)}]{Dixon:2009uk}%
  \BibitemOpen
  \bibfield  {author} {\bibinfo {author} {\bibfnamefont {L.~J.}\ \bibnamefont
  {Dixon}}\ and\ \bibinfo {author} {\bibfnamefont {Y.}~\bibnamefont
  {Sofianatos}},\ }\href {\doibase 10.1088/1126-6708/2009/08/058} {\bibfield
  {journal} {\bibinfo  {journal} {JHEP}\ }\textbf {\bibinfo {volume} {08}},\
  \bibinfo {pages} {058} (\bibinfo {year} {2009})},\ \Eprint
  {http://arxiv.org/abs/0906.0008} {arXiv:0906.0008 [hep-ph]} \BibitemShut
  {NoStop}%
\bibitem [{\citenamefont {Badger}\ \emph {et~al.}(2010)\citenamefont {Badger},
  \citenamefont {Nigel~Glover}, \citenamefont {Mastrolia},\ and\ \citenamefont
  {Williams}}]{Badger:2009hw}%
  \BibitemOpen
  \bibfield  {author} {\bibinfo {author} {\bibfnamefont {S.}~\bibnamefont
  {Badger}}, \bibinfo {author} {\bibfnamefont {E.~W.}\ \bibnamefont
  {Nigel~Glover}}, \bibinfo {author} {\bibfnamefont {P.}~\bibnamefont
  {Mastrolia}}, \ and\ \bibinfo {author} {\bibfnamefont {C.}~\bibnamefont
  {Williams}},\ }\href {\doibase 10.1007/JHEP01(2010)036} {\bibfield  {journal}
  {\bibinfo  {journal} {JHEP}\ }\textbf {\bibinfo {volume} {01}},\ \bibinfo
  {pages} {036} (\bibinfo {year} {2010})},\ \Eprint
  {http://arxiv.org/abs/0909.4475} {arXiv:0909.4475 [hep-ph]} \BibitemShut
  {NoStop}%
\bibitem [{\citenamefont {Badger}\ \emph {et~al.}(2009)\citenamefont {Badger},
  \citenamefont {Campbell}, \citenamefont {Ellis},\ and\ \citenamefont
  {Williams}}]{Badger:2009vh}%
  \BibitemOpen
  \bibfield  {author} {\bibinfo {author} {\bibfnamefont {S.}~\bibnamefont
  {Badger}}, \bibinfo {author} {\bibfnamefont {J.~M.}\ \bibnamefont
  {Campbell}}, \bibinfo {author} {\bibfnamefont {R.~K.}\ \bibnamefont {Ellis}},
  \ and\ \bibinfo {author} {\bibfnamefont {C.}~\bibnamefont {Williams}},\
  }\href {\doibase 10.1088/1126-6708/2009/12/035} {\bibfield  {journal}
  {\bibinfo  {journal} {JHEP}\ }\textbf {\bibinfo {volume} {12}},\ \bibinfo
  {pages} {035} (\bibinfo {year} {2009})},\ \Eprint
  {http://arxiv.org/abs/0910.4481} {arXiv:0910.4481 [hep-ph]} \BibitemShut
  {NoStop}%
\end{thebibliography}

\onecolumngrid
\newpage
\appendix

\section*{Supplemental material}

In this supplemental material, we first describe the subtraction of ultraviolet (UV) and infrared (IR) divergences. Then we give the explicit maximal transcendentality parts of the remainder functions for the form factors that involve $q\bar q g$ external states.

\subsection{Renormalization and IR subtraction}
\label{app:divergence-structure}
The bare form factors contain ultraviolet (UV) and infrared (IR) divergences. In this appendix we describe explicit how these divergences are removed which leads to the finite remainder functions that we consider in the main text. We apply dimensional regularization ($D=4-2\epsilon$) and use the $\overline{\textrm{MS}}$ scheme. Most formulas can be also found in \cite{Gehrmann:2011aa}. Here we present the formula depending explicitly on the quadratic Casimirs $C_A$ and $C_F$ rather than taking $N_c$ expansion.

The bare form factor is expanded as
\begin{align}
{\cal F}_{\rm b} = g_0^x \left[ {\cal F}_{\rm b}^{(0)} + {\alpha_0 \over 4\pi} {\cal F}_{\rm b}^{(1)} + \Big( {\alpha_0 \over 4\pi} \Big)^2 {\cal F}_{\rm b}^{(2)} + {\cal O}(\alpha_0^3) \right] \,,
\end{align}
where $g_0 = g_{\textrm{\tiny YM}}$ is the bare gauge coupling and $\alpha_0 = \frac{g_0^2}{4\pi}$. We pull out the coupling $g_0^x$ in the tree form factor which depends on the number of external legs. 

To remove the UV divergences, we preform renormalization for both the coupling constant and the local operator. 
We express the bare coupling $\alpha_0$ in terms of the renormalized coupling $\alpha_s =\alpha_s(\mu^2) = \frac{g_s(\mu^2)^2}{4\pi}$, evaluated at the renormalization scale $\mu^2$, as
\begin{align}
\alpha_0 & = \alpha_s  S_\epsilon^{-1} {\mu^{2\epsilon} \over \mu_0^{2\epsilon}}  \Big[ 1 - {\beta_0 \over \epsilon} {\alpha_s \over 4\pi} + \Big( {\beta_0^2 \over \epsilon^2} - {\beta_1 \over 2 \epsilon} \Big) \Big({\alpha_s \over 4\pi}\Big)^2 + {\cal O}(\alpha_s^3) \Big] \,, 
\end{align}
where $S_\epsilon = (4\pi e^{- \gamma_{\text{E}}})^\epsilon$, due to the use of ${\overline {\rm MS}}$ scheme, and $\mu_0^2$ is the scale introduced to keep gauge coupling dimensionless in the bare Lagrangian. 
The first two coefficients of the $\beta$ function are
\begin{align}
\beta_0 = {11 C_A \over 3} - {2 n_f \over 3} \,, \qquad \beta_1 = {34 C_A^2 \over 3} - {10 C_A n_f \over 3} - 2 C_F n_f \,,
\end{align}
where $n_f$ is the number of fermion flavors. 
In $SU(N_c)$, the quadratic Casimirs in the adjoint and fundamental representations respectively are
\begin{align}
C_A = N_c \,, \qquad C_F = \frac{N_c^2 -1}{2N_c} \,.
\end{align}
We renormalize the operator by introducing the renormalization constant  $Z$ for the operator
\begin{equation}
Z = 1 + \sum_{l=1}^\infty \Big({\alpha_s \over 4\pi}\Big)^l Z^{(l)}  \,.
\label{eq:Z_def_expand}
\end{equation}

Expanding the renormalized form factor as 
\begin{align}
{\cal F} = g_s^x \, S_\epsilon^{-x/2} \sum_{l=0}^\infty \Big( {\alpha_s \over 4\pi} \Big)^l {\cal F}^{(l)}  \,,
\end{align}
we have the relations between the renormalized components ${\cal F}^{(l)}$ and the bare ones ${\cal F}_{\rm b}^{(l)}$ as
\begin{align}
{\cal F}^{(0)} & = {\cal F}_{\rm b}^{(0)} \,, \\
{\cal F}^{(1)} & = S_\epsilon^{-1} {\cal F}_{\rm b}^{(1)} +  \Big( Z^{(1)} - {x\over2} {\beta_0 \over \epsilon} \Big) {\cal F}_{\rm b}^{(0)}  \,, \\
{\cal F}^{(2)} & = S_\epsilon^{-2} {\cal F}_{\rm b}^{(2)} + {S_\epsilon^{-1}} \Big[ Z^{(1)} - \Big(1+{x\over2} \Big) {\beta_0 \over \epsilon} \Big] {\cal F}_{\rm b}^{(1)} \nonumber\\
& +  \Big[ Z^{(2)} - {x\over2} {\beta_0 \over \epsilon} Z^{(1)} + {x^2+2x\over8} {\beta_0^2 \over \epsilon^2} - {x\over4} {\beta_1 \over \epsilon} \Big] {\cal F}_{\rm b}^{(0)} \,.
\end{align}

The renormalized form factor contains only IR divergences, which take following universal structure: 
\begin{align}
{\cal F}^{(1)} &= I^{(1)}(\epsilon) {\cal F}^{(0)} + {\cal F}^{(1),{\rm fin}} + {\cal O}(\epsilon) \,,  \\
{\cal F}^{(2)} &= I^{(2)}(\epsilon) {\cal F}^{(0)} +  I^{(1)}(\epsilon) {\cal F}^{(1)} + {\cal F}^{(2),{\rm fin}} + {\cal O}(\epsilon)   \,.
\end{align}

At one-loop, for the form factor with three external gluons, we have
\begin{align}
I_{3g}^{(1)}(\epsilon) = - {e^{\gamma_E \epsilon} \over \Gamma(1-\epsilon)} \bigg( \frac{C_A}{\epsilon^2} + \frac{\beta_0}{2 \epsilon} \bigg) \sum_{i=1}^3 (-{s_{i,i+1}\over\mu^2} )^{-\epsilon} \,, 
\end{align}
while for the case with two external quarks plus one gluon, we have
\begin{align}
I_{ffg}^{(1)}(\epsilon) = - {e^{\gamma_E \epsilon} \over \Gamma(1-\epsilon)} \bigg[ \bigg( \frac{C_A}{\epsilon^2} + \frac{3C_A}{4\epsilon} + \frac{\beta_0}{4 \epsilon} \bigg) \Big( (-{s_{13}\over\mu^2} )^{-\epsilon} + (-{s_{23}\over\mu^2} )^{-\epsilon} \Big) - (C_A - 2 C_F) \bigg( \frac{1}{\epsilon^2} + \frac{3}{2\epsilon} \bigg) (-{s_{12}\over\mu^2} )^{-\epsilon}  \bigg] \,.
\end{align}

At two-loop, we have the general structure that, for $n$ external gluons and $m$ external quarks (or anti-quarks),
\begin{align}
I^{(2)}(\epsilon) &= - {1\over2} \big[ I^{(1)}(\epsilon) \big]^2   -  {\beta_0 \over \epsilon} I^{(1)}(\epsilon)  + {e^{-\gamma_E \epsilon} \Gamma(1-2\epsilon) \over \Gamma(1-\epsilon)} \left[ \frac{\beta_0}{\epsilon} + {\cal K} \right] I^{(1)}(2\epsilon) + {e^{\gamma_E \epsilon} \over \epsilon \Gamma(1-\epsilon)} \Big( n {\cal H}_{\Omega,g}^{(2)} + m {\cal H}_{\Omega,q}^{(2)} \Big) \,, \nonumber
\end{align}
where
\begin{equation}
{\cal K} = \left({67\over9} - {\pi^2\over3}\right) C_A - {10\over9}n_f \,,
\end{equation}
and
\begin{align}
{\cal H}_{\Omega,g}^{(2)}  = &  \left( \frac{\zeta_3}{2} + {5\over12} + {11\pi^2 \over 144} \right)C_A^2 + {5 n_f^2 \over 27} - \left( {\pi^2 \over 72} + {89 \over 108} \right) C_A n_f - {n_f(C_A - 2 C_F)\over 4}  \,, \\
{\cal H}_{\Omega,q}^{(2)}  = &  \left( \frac{7\zeta_3}{4} + {409\over864} - {11\pi^2 \over 96} \right)C_A^2 - \left( \frac{\zeta_3}{4} + {41\over108} + {\pi^2 \over 96} \right) C_A(C_A - 2 C_F)  \nonumber\\
& - \left( \frac{3\zeta_3}{2} + {3\over 32} - {\pi^2 \over 8} \right) {(C_A - 2 C_F)^2} + \left( {\pi^2 \over 48} - {25 \over 216} \right) { 2 C_F n_f}  \,.
\end{align}

\subsection{Maximally Transcendental Remainder of ${\cal R}^{(2)}_{{\cal O};4}(1^q, 2^{\bar q}, 3^\pm)$}
\label{app:simp-quark-remainder}

In this appendix, we give the maximal transcendentality part of the remainder function for the form factors with length-2 operators: ${\cal O}_0$=${\rm tr}(F^2)$, ${\rm tr}(\bar\psi\psi)$, and length-3 operator ${\cal O}_4\sim F_{\mu\nu}D^\mu(\bar\psi\gamma^\nu\psi)$, respectively.  The Mathematica readable format is also provided with the submission of this article.

The first case, ${\cal R}^{(2)}_{{\cal O}_0;4}(1^q, 2^{\bar q}, 3^\pm)$, has been already given in \cite{Gehrmann:2011aa}. We have reproduced this result by an independent computation. We expand the result in three terms associated with following three color factors:
\begin{align}
{\cal R}^{(2)}_{{\cal O}_0;4}(1^q, 2^{\bar q}, 3^\pm)  = C_A^2 {\cal R}^{(2), C_A^2}_{{\cal O}_0;4}(1^q, 2^{\bar q}, 3^\pm) + C_A C_F{\cal R}^{(2),C_A C_F}_{{\cal O}_0;4}(1^q, 2^{\bar q}, 3^\pm) + C_F^2 {\cal R}^{(2), C_F^2}_{{\cal O}_0;4}(1^q, 2^{\bar q}, 3^\pm) \,.
\end{align}
Each term is a non-trivial functions in terms of  2d Harmonic polylogarithms. We express the result in a basis of  2d Harmonic polylogarithms \cite{Gehrmann:2001jv}: $G({\vec m}(z); y)$ and $G({\vec m}; z)$, where $y=v={s_{23}\over q^2}, z=w={s_{13}\over q^2}$. Explicitly, they are:
\begin{dmath}
{\cal R}^{(2), C_A^2}_{{\cal O}_0;4} =
\frac{4}{3} G(0,z) G(1,z)^3-\frac{3}{2} G(0,z)^2 G(1,z)^2+2 G(-z,y)^2 G(1,z)^2-2 G(0,z) G(-z,y) G(1,z)^2-\frac{4}{3}
   G(-z,y)^3 G(1,z)-4 G(0,0,-z,y) G(1,z)+2 G(-z,0,1-z,y) G(1,z)-3 G(0,1,z)^2+G(0,y)^2 \left(-\frac{3}{2} G(0,z)^2+G(1,z)
   G(0,z)+\frac{1}{2} G(1,z)^2+\frac{1}{2} G(1-z,y)^2+G(1,y) (2 G(0,z)-2 G(1,z)-2 G(1-z,y))+(G(0,z)+G(1,z))
   G(1-z,y)\right)+G(1-z,y)^2 \left(\frac{1}{2} G(0,z)^2+2 G(1,z) G(0,z)-2 G(1,z) G(-z,y)\right)+G(1-z,y) \left(-G(1,z)
   G(0,z)^2+4 G(1,z)^2 G(0,z)-2 G(1,z) G(-z,y) G(0,z)+4 G(1,z) G(-z,y)^2\right)+G(0,y) \left(3 G(1,z) G(0,z)^2-2 G(1,z)^2
   G(0,z)-G(1-z,y)^2 G(0,z)+\left(2 G(0,z) G(1,z)-2 G(1,z)^2\right) G(-z,y)+G(1-z,y) \left(G(0,z)^2-4 G(1,z) G(0,z)-2
   G(1,z) G(-z,y)\right)\right)+G(1,y) \left(G(1-z,y) (2 G(1,z) G(-z,y)-2 G(0,z) G(1,z))-G(0,z) G(1,z)^2\right)+\left(-3
   G(1,z)^2+8 G(-z,y) G(1,z)-4 G(1-z,y)^2-4 G(-z,y)^2+G(0,y) (-4 G(0,z)-2 G(1,z))+G(1,y) (2 G(0,z)+4 G(1,z)+4
   G(1-z,y))+G(1-z,y) (8 G(-z,y)-8 G(1,z))\right) G(0,1,z)+(-2 G(0,y) G(0,z)-2 G(1,y) G(0,z)+4 G(1,z) G(0,z)-2 G(1,z)
   G(-z,y)-2 G(0,1,z)) G(0,1-z,y)+(2 G(0,y) G(1,z)-2 G(0,z) G(1,z)+2 G(1,y) G(1,z)) G(0,-z,y)+G(0,1,y) (2 G(0,z)
   G(1-z,y)+G(0,y) (-4 G(0,z)+4 G(1,z)+4 G(1-z,y))-2 G(1,z) G(-z,y)-2 G(-z,1-z,y))+G(1-z,1,y) (2 G(0,z) G(1,z)-2 G(-z,y)
   G(1,z)+G(0,y) (-2 G(0,z)+4 G(1,z)+4 G(1-z,y))-4 G(0,1,z)-2 G(-z,1-z,y))+\left(-2 G(1-z,y)^2+(4 G(1,z)-2 G(0,z))
   G(1-z,y)-2 G(0,z) G(1,z)+G(0,y) (2 G(0,z)-2 G(1,z)-2 G(1-z,y))+G(1,y) (2 G(1-z,y)-2 G(1,z))\right) G(-z,1-z,y)+(4
   G(0,z)-4 G(1,z)-4 G(1-z,y)) G(0,0,1,y)+(6 G(0,y)+2 G(0,z)-6 G(1,y)+10 G(1,z)+8 G(1-z,y)-8 G(-z,y)) G(0,0,1,z)+4 G(0,z)
   G(0,0,1-z,y)+(6 G(0,y)+4 G(0,z)-6 G(1,y)+4 G(1,z)+8 G(1-z,y)-8 G(-z,y)) G(0,1,1,z)+(2 G(0,z)-4 G(1,z)-4 G(1-z,y))
   G(0,1-z,1,y)+4 G(0,z) G(0,1-z,1-z,y)+(2 G(0,y)-2 G(0,z)+2 G(1,y)+2 G(1,z)) G(0,-z,1-z,y)+(-2 G(0,y)-4 G(1,z)-4 G(1-z,y))
   G(1-z,0,1,y)-4 G(0,y) G(1-z,1-z,1,y)+(2 G(1,z)+2 G(1-z,y)) G(-z,0,1,y)+(2 G(1,z)+2 G(1-z,y)) G(-z,1-z,1,y)+(-4 G(1,y)-4
   G(1,z)+8 G(1-z,y)) G(-z,1-z,1-z,y)+8 G(1-z,y) G(-z,-z,1-z,y)-12 G(0,0,0,1,z)-8 G(0,0,1,1,z)-4 G(0,0,-z,1-z,y)-2
   G(0,1,1,1,z)+2 G(0,1-z,0,1,y)+4 G(0,1-z,1-z,1,y)-2 G(0,1-z,-z,1-z,y)+4 G(1-z,0,0,1,y)+4 G(1-z,0,1-z,1,y)+4
   G(1-z,1-z,0,1,y)-2 G(1-z,-z,0,1,y)-2 G(1-z,-z,1-z,1,y)-12 G(-z,1-z,1-z,1-z,y)-8 G(-z,-z,1-z,1-z,y)-8
   G(-z,-z,-z,1-z,y)+\left(-3 G(1,z)^2+3 G(0,z) G(1,z)+2 G(-z,y) G(1,z)-4 G(1-z,y)^2+(3 G(0,z)-8 G(1,z)) G(1-z,y)+G(1,y)
   (-2 G(0,z)+2 G(1,z)+2 G(1-z,y))+G(0,y) (-G(0,z)+G(1,z)+3 G(1-z,y))-2 G(0,1-z,y)-2 G(1-z,1,y)+2 G(-z,1-z,y)\right) \zeta
   (2)+\frac{119 \zeta (4)}{8}+(-7 G(0,y)-7 G(0,z)+6 G(1,y)+7 G(1,z)+G(1-z,y)) \zeta (3) \,,
\end{dmath}
\begin{dmath}
{\cal R}^{(2), C_A C_F}_{{\cal O}_0;4} =
-\frac{4}{3} G(0,z) G(1,z)^3+3 G(0,z)^2 G(1,z)^2-8 G(0,0,-z,y) G(1,z)-10 G(-z,0,1-z,y) G(1,z)-2 G(0,1,y)^2+11
   G(0,1,z)^2+\left(2 G(0,z)^2 G(1,z)-2 G(0,z) G(1,z)^2\right) G(1-z,y)+G(0,y)^2 \left(3 G(0,z)^2-2 G(1,z) G(0,z)-2
   G(1-z,y) G(0,z)+G(1,y) (-4 G(0,z)+4 G(1,z)+4 G(1-z,y))\right)+G(0,y) \left(-6 G(1,z) G(0,z)^2+3 G(1,z)^2 G(0,z)+\left(4
   G(0,z) G(1,z)-2 G(0,z)^2\right) G(1-z,y)+G(1,y) (2 G(0,z) G(1,z)+2 G(0,z) G(1-z,y))\right)+G(1,y) \left(3 G(0,z)
   G(1,z)^2+G(1-z,y) (6 G(0,z) G(1,z)-10 G(1,z) G(-z,y))\right)+\left(-2 G(0,y)^2+(8 G(0,z)-4 G(1,y)+10 G(1,z)+4 G(1-z,y))
   G(0,y)-5 G(1,z)^2-8 G(0,z) G(1,z)+G(1,y) (-6 G(0,z)-16 G(1,z)-16 G(1-z,y))+(-4 G(0,z)-4 G(1,z)) G(1-z,y)\right)
   G(0,1,z)+\left(-2 G(0,z)^2+2 G(0,y) G(0,z)+6 G(1,y) G(0,z)+8 G(1-z,y) G(0,z)+10 G(1,z) G(-z,y)+2 G(0,1,z)\right)
   G(0,1-z,y)+(2 G(0,y) G(1,z)-2 G(0,z) G(1,z)-10 G(1,y) G(1,z)) G(0,-z,y)+G(1,y) (10 G(1,z)-10 G(1-z,y))
   G(-z,1-z,y)+G(1-z,1,y) \left(-2 G(0,y)^2+(2 G(0,z)-4 G(1,y)-8 G(1,z)-8 G(1-z,y)) G(0,y)-6 G(0,z) G(1,z)+10 G(1,z)
   G(-z,y)+16 G(0,1,z)+10 G(-z,1-z,y)\right)+G(0,1,y) (-2 G(0,z) G(1,z)+10 G(-z,y) G(1,z)+G(0,y) (6 G(0,z)-4 G(1,y)-6
   G(1,z)-6 G(1-z,y))-8 G(0,z) G(1-z,y)+4 G(0,1,z)-8 G(1-z,1,y)+10 G(-z,1-z,y))+(4 G(0,y)-4 G(0,z)+8 G(1,y)+4 G(1,z)+4
   G(1-z,y)) G(0,0,1,y)+(-10 G(0,y)+6 G(0,z)+22 G(1,y)+2 G(1,z)+4 G(1-z,y)) G(0,0,1,z)-4 G(0,z) G(0,0,1-z,y)+8 G(0,y)
   G(0,1,1,y)+(-26 G(0,y)+26 G(1,y)+28 G(1,z)+12 G(1-z,y)) G(0,1,1,z)+(-4 G(0,y)-2 G(0,z)+4 G(1,y)+8 G(1,z)+8 G(1-z,y))
   G(0,1-z,1,y)-16 G(0,z) G(0,1-z,1-z,y)+(2 G(0,y)-2 G(0,z)-10 G(1,y)-10 G(1,z)) G(0,-z,1-z,y)+(2 G(0,y)+4 G(0,z)+4
   G(1,y)+8 G(1,z)+8 G(1-z,y)) G(1-z,0,1,y)+12 G(0,y) G(1-z,1,1,y)+16 G(0,y) G(1-z,1-z,1,y)+(-10 G(1,z)-10 G(1-z,y))
   G(-z,0,1,y)+(-10 G(1,z)-10 G(1-z,y)) G(-z,1-z,1,y)+20 G(1,y) G(-z,1-z,1-z,y)-12 G(0,0,0,1,y)-20 G(0,0,0,1,z)-8
   G(0,0,1,1,y)-28 G(0,0,1,1,z)+12 G(0,0,1-z,1,y)-8 G(0,0,-z,1-z,y)-46 G(0,1,1,1,z)+8 G(0,1,1-z,1,y)+6 G(0,1-z,0,1,y)+4
   G(0,1-z,1,1,y)-16 G(0,1-z,1-z,1,y)+10 G(0,1-z,-z,1-z,y)+4 G(1-z,0,1,1,y)-16 G(1-z,0,1-z,1,y)-16 G(1-z,1-z,0,1,y)+10
   G(1-z,-z,0,1,y)+10 G(1-z,-z,1-z,1,y)+\left(-2 G(0,y)^2+(-8 G(0,z)+2 G(1,y)+10 G(1,z)+4 G(1-z,y)) G(0,y)-2 G(0,z)^2-5
   G(1,z)^2+6 G(0,z) G(1,z)+G(1,y) (6 G(0,z)-6 G(1,z)-6 G(1-z,y))+(4 G(0,z)-4 G(1,z)) G(1-z,y)-2 G(0,1,y)+2 G(0,1-z,y)+2
   G(1-z,1,y)\right) \zeta (2)-\frac{311 \zeta (4)}{4}+(22 G(0,y)+22 G(0,z)-18 G(1,y)-52 G(1,z)-34 G(1-z,y))
   \zeta (3) \,,
\end{dmath}
\begin{dmath}
{\cal R}^{(2), C_F^2}_{{\cal O}_0;4} =
\left(-G(0,z)^2+2 G(1,z) G(0,z)+2 G(1-z,y) G(0,z)+2 G(1,y)^2+G(1,y) (-4 G(1,z)-4 G(1-z,y))\right)
   G(0,y)^2+\left(-\frac{4}{3} G(1,y)^3-2 G(0,z) G(1,z)^2+2 G(0,z)^2 G(1,z)+\left(2 G(0,z)^2-4 G(0,z) G(1,z)\right)
   G(1-z,y)\right) G(0,y)-12 G(1-z,1,1,y) G(0,y)-16 G(1-z,1-z,1,y) G(0,y)-G(0,z)^2 G(1,z)^2+2 G(0,1,y)^2-8
   G(0,1,z)^2+\left(2 G(0,z) G(1,z)^2-2 G(0,z)^2 G(1,z)\right) G(1-z,y)+G(1,y) \left(G(1-z,y) (8 G(1,z) G(-z,y)-4 G(0,z)
   G(1,z))-2 G(0,z) G(1,z)^2\right)+\left(2 G(0,y)^2+(-4 G(0,z)+4 G(1,y)-8 G(1,z)-4 G(1-z,y)) G(0,y)+8 G(1,z)^2+8 G(0,z)
   G(1,z)+(4 G(0,z)+4 G(1,z)) G(1-z,y)+G(1,y) (4 G(0,z)+12 G(1,z)+12 G(1-z,y))\right) G(0,1,z)+\left(2 G(0,z)^2-4 G(0,y)
   G(0,z)-4 G(1,y) G(0,z)-8 G(1-z,y) G(0,z)-8 G(1,z) G(-z,y)\right) G(0,1-z,y)+8 G(1,y) G(1,z) G(0,-z,y)+G(1-z,1,y) \left(2
   G(0,y)^2+(-4 G(0,z)+4 G(1,y)+8 G(1,z)+8 G(1-z,y)) G(0,y)+4 G(0,z) G(1,z)-8 G(1,z) G(-z,y)-12 G(0,1,z)-8
   G(-z,1-z,y)\right)+G(0,1,y) \left(4 G(1,y)^2+4 G(0,z) G(1-z,y)+G(0,y) (4 G(1,y)+4 G(1,z)+4 G(1-z,y))-8 G(1,z) G(-z,y)-4
   G(0,1,z)+8 G(1-z,1,y)-8 G(-z,1-z,y)\right)+G(1,y) (8 G(1-z,y)-8 G(1,z)) G(-z,1-z,y)+(-8 G(0,y)-16 G(1,y)) G(0,0,1,y)+(4
   G(0,y)-8 G(0,z)-16 G(1,y)-12 G(1,z)-4 G(1-z,y)) G(0,0,1,z)+4 G(0,z) G(0,0,1-z,y)+8 G(1,z) G(0,0,-z,y)+(-8 G(0,y)-8
   G(1,y)) G(0,1,1,y)+(20 G(0,y)-4 G(0,z)-20 G(1,y)-32 G(1,z)-12 G(1-z,y)) G(0,1,1,z)+(4 G(0,y)+4 G(0,z)-4 G(1,y)-8
   G(1,z)-8 G(1-z,y)) G(0,1-z,1,y)+16 G(0,z) G(0,1-z,1-z,y)+(8 G(1,y)+8 G(1,z)) G(0,-z,1-z,y)+(-4 G(1,y)-8 G(1,z)-8
   G(1-z,y)) G(1-z,0,1,y)+(8 G(1,z)+8 G(1-z,y)) G(-z,0,1,y)+8 G(1,z) G(-z,0,1-z,y)+(8 G(1,z)+8 G(1-z,y)) G(-z,1-z,1,y)-16
   G(1,y) G(-z,1-z,1-z,y)+24 G(0,0,0,1,y)+32 G(0,0,0,1,z)+16 G(0,0,1,1,y)+36 G(0,0,1,1,z)-12 G(0,0,1-z,1,y)+8
   G(0,0,-z,1-z,y)+8 G(0,1,1,1,y)+48 G(0,1,1,1,z)-8 G(0,1,1-z,1,y)-8 G(0,1-z,0,1,y)-4 G(0,1-z,1,1,y)+16 G(0,1-z,1-z,1,y)-8
   G(0,1-z,-z,1-z,y)-4 G(1-z,0,0,1,y)-4 G(1-z,0,1,1,y)+16 G(1-z,0,1-z,1,y)+16 G(1-z,1-z,0,1,y)-8 G(1-z,-z,0,1,y)-8
   G(1-z,-z,1-z,1,y)+\left(2 G(0,y)^2+(12 G(0,z)-8 G(1,y)-8 G(1,z)-4 G(1-z,y)) G(0,y)+2 G(0,z)^2+4 G(1,y)^2+8 G(1,z)^2-12
   G(0,z) G(1,z)+(4 G(1,z)-4 G(0,z)) G(1-z,y)+G(1,y) (-4 G(0,z)+4 G(1,z)+4 G(1-z,y))\right) \zeta (2)+69 \zeta (4)+(-16
   G(0,y)-16 G(0,z)+12 G(1,y)+44 G(1,z)+32 G(1-z,y)) \zeta (3) \,.
\end{dmath}
Although each terms is very complicated, the sum of the three terms (or equivalently, by changing $C_F$ to be $C_A$) is remarkably simple as given in \eqref{eq:L2-quarksFF-relation}, which is expressed in terms of only classical polylogarithms.

Next, we give the result of ${\cal R}^{(2)}_{\bar\psi \psi;4}(1^q, 2^{\bar q}, 3^\pm)$. It has the same color structure as previous case, and we also express the result in the same basis of 2d Harmonic polylogarithms: 
\begin{align}
{\cal R}^{(2)}_{\bar\psi \psi;4}(1^q, 2^{\bar q}, 3^\pm) =  C_A^2 {\cal R}^{(2), C_A^2}_{\bar\psi \psi;4}(1^q, 2^{\bar q}, 3^\pm) + C_A C_F {\cal R}^{(2),C_A C_F}_{\bar\psi \psi;4}(1^q, 2^{\bar q}, 3^\pm)  + C_F^2 {\cal R}^{(2), C_F^2}_{\bar\psi \psi;4}(1^q, 2^{\bar q}, 3^\pm) \,,
\end{align}
where
\begin{dmath}
 {\cal R}^{(2), C_A^2}_{\bar\psi \psi;4}  =
 -G(0,z) G(1,z)^3+\frac{1}{2} G(0,z)^2 G(1,z)^2+\frac{8}{3} G(-z,y)^3 G(1,z)-2 G(0,z) G(-z,y)^2 G(1,z)+8 G(0,0,-z,y)
   G(1,z)+2 G(-z,0,1-z,y) G(1,z)+G(0,1,z)^2+4 G(-z,1-z,y)^2+G(0,y)^2 \left(\frac{1}{2} G(0,z)^2-G(1,z) G(0,z)+\frac{1}{2}
   G(1,z)^2+\frac{1}{2} G(1-z,y)^2+(G(1,z)-G(0,z)) G(1-z,y)\right)+G(1-z,y)^2 \left(\frac{1}{2} G(0,z)^2-2 G(1,z) G(0,z)-2
   G(1,z) G(-z,y)\right)+G(1-z,y) \left(G(1,z) G(0,z)^2-3 G(1,z)^2 G(0,z)+2 G(1,z) G(-z,y) G(0,z)-4 G(1,z)
   G(-z,y)^2\right)+G(1,y) \left(G(1-z,y) (2 G(1,z) G(-z,y)-2 G(0,z) G(1,z))-G(0,z) G(1,z)^2\right)+G(0,y) \left(-G(1,z)
   G(0,z)^2+G(1,z)^2 G(0,z)+G(1-z,y)^2 G(0,z)+2 G(1,z) G(-z,y) G(0,z)-2 G(1,z) G(-z,y)^2+G(1-z,y) \left(-G(0,z)^2+2 G(1,z)
   G(0,z)+2 G(1,z) G(-z,y)\right)\right)+\left(G(1,z)^2+4 G(0,z) G(1,z)+8 G(-z,y)^2+G(1,y) (2 G(0,z)+4 G(1,z)+4
   G(1-z,y))+G(1-z,y) (4 G(0,z)-8 G(-z,y))+G(0,y) (2 G(0,z)-2 G(1,z)-4 G(-z,y))-4 G(0,z) G(-z,y)\right) G(0,1,z)+(-2 G(0,z)
   G(1,y)-2 G(0,z) G(1,z)-2 G(0,z) G(1-z,y)-2 G(1,z) G(-z,y)-2 G(0,1,z)) G(0,1-z,y)+(-4 G(0,y) G(1,z)+4 G(0,z) G(1,z)+2
   G(1,y) G(1,z)) G(0,-z,y)+G(0,1,y) (2 G(0,z) G(1-z,y)-2 G(1,z) G(-z,y)-2 G(-z,1-z,y))+G(1-z,1,y) (2 G(0,z) G(1,z)-2
   G(-z,y) G(1,z)+G(0,y) (-2 G(1,z)-2 G(1-z,y))-4 G(0,1,z)-2 G(-z,1-z,y))+\left(-2 G(1-z,y)^2+(2 G(0,z)+4 G(1,z))
   G(1-z,y)-2 G(0,z) G(1,z)+G(1,y) (2 G(1-z,y)-2 G(1,z))+G(0,y) (2 G(0,z)-2 G(1,z)+2 G(1-z,y))+8 G(1,z) G(-z,y)+8
   G(0,1,z)\right) G(-z,1-z,y)+(-8 G(0,z)-6 G(1,y)-10 G(1,z)-12 G(1-z,y)+16 G(-z,y)) G(0,0,1,z)+(-4 G(0,z)-6 G(1,y)+2
   G(1,z)+6 G(1-z,y)) G(0,1,1,z)+(4 G(0,y)+2 G(1,z)+2 G(1-z,y)) G(0,1-z,1,y)+(-4 G(0,y)+4 G(0,z)+2 G(1,y)+2 G(1,z))
   G(0,-z,1-z,y)+(2 G(0,y)-2 G(0,z)+2 G(1,z)+2 G(1-z,y)) G(1-z,0,1,y)+(2 G(1,z)+2 G(1-z,y)) G(-z,0,1,y)+(2 G(1,z)+2
   G(1-z,y)) G(-z,1-z,1,y)+(-4 G(0,y)-4 G(0,z)-4 G(1,y)-4 G(1,z)+8 G(1-z,y)) G(-z,1-z,1-z,y)+(-4 G(0,y)-4 G(0,z)-8 G(1,z)-8
   G(1-z,y)) G(-z,-z,1-z,y)+24 G(0,0,0,1,z)+8 G(0,0,1,1,z)-8 G(0,0,1-z,1,y)+8 G(0,0,-z,1-z,y)-6 G(0,1,1,1,z)-6
   G(0,1-z,0,1,y)-2 G(0,1-z,-z,1-z,y)-4 G(1-z,0,0,1,y)-2 G(1-z,-z,0,1,y)-2 G(1-z,-z,1-z,1,y)-12 G(-z,1-z,1-z,1-z,y)+16
   G(-z,-z,-z,1-z,y)+\left(G(1,z)^2-G(0,z) G(1,z)+2 G(-z,y) G(1,z)+G(0,y) (3 G(0,z)-3 G(1,z)-G(1-z,y))-G(0,z)
   G(1-z,y)+G(1,y) (-2 G(0,z)+2 G(1,z)+2 G(1-z,y))-2 G(0,1-z,y)-2 G(1-z,1,y)+2 G(-z,1-z,y)\right) \zeta (2)+\frac{39 \zeta
   (4)}{8}+(-G(0,y)-G(0,z)+6 G(1,y)-5 G(1,z)-11 G(1-z,y)) \zeta (3) \,,
\end{dmath}
\begin{dmath}
{\cal R}^{(2), C_A C_F}_{\bar\psi \psi;4} = 
\frac{7}{3} G(0,z) G(1,z)^3-2 G(0,z) G(-z,y) G(1,z)^2-\frac{20}{3} G(-z,y)^3 G(1,z)-20 G(0,0,-z,y) G(1,z)+\left(6 G(0,z)
   G(1,z)-2 G(1,z)^2\right) G(-z,y)^2+2 G(0,1,y)^2+G(0,1,z)^2-12 G(-z,1-z,y)^2+G(0,y)^2 \left(-2 G(1,z)^2+2 G(0,z) G(1,z)-2
   G(1-z,y)^2+(2 G(0,z)-4 G(1,z)) G(1-z,y)+G(1,y) (-2 G(0,z)+2 G(1,z)+2 G(1-z,y))\right)+G(1-z,y)^2 \left(-2 G(0,z)^2+4
   G(1,z) G(0,z)+8 G(1,z) G(-z,y)\right)+G(1-z,y) \left(-2 G(1,z) G(0,z)^2+9 G(1,z)^2 G(0,z)-4 G(1,z) G(-z,y) G(0,z)+8
   G(1,z) G(-z,y)^2\right)+G(1,y) \left(G(0,z) G(1,z)^2+G(1-z,y) (2 G(0,z) G(1,z)-6 G(1,z) G(-z,y))\right)+G(0,y) \left(-2
   G(0,z) G(1,z)^2+2 G(-z,y) G(1,z)^2+6 G(-z,y)^2 G(1,z)-2 G(0,z) G(1-z,y)^2+G(1,y) (2 G(0,z) G(1,z)+2 G(0,z)
   G(1-z,y))+G(1-z,y) \left(2 G(0,z)^2-6 G(1,z) G(0,z)-4 G(1,z) G(-z,y)\right)\right)+\left(-G(1,z)^2-4 G(0,z) G(1,z)+4
   G(1-z,y)^2-20 G(-z,y)^2+G(1,y) (-2 G(0,z)-8 G(1,z)-8 G(1-z,y))+4 G(0,z) G(-z,y)+G(0,y) (-2 G(0,z)-4 G(1,y)+6 G(1,z)+4
   G(1-z,y)+4 G(-z,y))+G(1-z,y) (16 G(-z,y)-8 G(0,z))\right) G(0,1,z)+(-4 G(0,y) G(0,z)+2 G(1,y) G(0,z)+14 G(1,z) G(0,z)+6
   G(1-z,y) G(0,z)+6 G(1,z) G(-z,y)-6 G(0,1,z)) G(0,1-z,y)+(8 G(0,y) G(1,z)-16 G(0,z) G(1,z)-6 G(1,y) G(1,z)+8 G(0,1,z))
   G(0,-z,y)+\left(8 G(1-z,y)^2+(-4 G(0,z)-16 G(1,z)) G(1-z,y)+G(1,y) (6 G(1,z)-6 G(1-z,y))+G(0,y) (8 G(1,z)-4 G(1-z,y))-24
   G(1,z) G(-z,y)-16 G(0,1,z)\right) G(-z,1-z,y)+G(1-z,1,y) (-2 G(0,z) G(1,z)+6 G(-z,y) G(1,z)+G(0,y) (-4 G(0,z)-4 G(1,y)+2
   G(1,z)+2 G(1-z,y))+8 G(0,1,z)+6 G(-z,1-z,y))+G(0,1,y) (-2 G(0,z) G(1,z)+6 G(-z,y) G(1,z)+G(0,y) (2 G(0,z)+4 G(1,y)-2
   G(1,z)-2 G(1-z,y))-4 G(0,z) G(1-z,y)+4 G(0,1,z)-12 G(1-z,1,y)+6 G(-z,1-z,y))+(-4 G(0,y)-8 G(1,y)) G(0,0,1,y)+(8
   G(0,z)+10 G(1,y)+6 G(1,z)+20 G(1-z,y)-24 G(-z,y)) G(0,0,1,z)+12 G(0,z) G(0,0,1-z,y)-8 G(0,y) G(0,1,1,y)+(-4 G(0,z)+14
   G(1,y)-10 G(1,z)-18 G(1-z,y)) G(0,1,1,z)+(-4 G(0,y)+4 G(0,z)+4 G(1,y)-2 G(1,z)-2 G(1-z,y)) G(0,1-z,1,y)-4 G(0,z)
   G(0,1-z,1-z,y)+(8 G(0,y)-16 G(0,z)-6 G(1,y)-6 G(1,z)) G(0,-z,1-z,y)+(-2 G(0,y)+6 G(0,z)+4 G(1,y)-2 G(1,z)-2 G(1-z,y))
   G(1-z,0,1,y)+16 G(0,y) G(1-z,1,1,y)+4 G(0,y) G(1-z,1-z,1,y)+(-8 G(0,y)-6 G(1,z)-6 G(1-z,y)) G(-z,0,1,y)+(-8 G(0,z)-6
   G(1,z)) G(-z,0,1-z,y)+(-8 G(0,y)-6 G(1,z)-6 G(1-z,y)) G(-z,1-z,1,y)+(12 G(0,y)+12 G(0,z)+12 G(1,y)+16 G(1,z)-32
   G(1-z,y)) G(-z,1-z,1-z,y)+(12 G(0,y)+12 G(0,z)+24 G(1,z)+16 G(1-z,y)) G(-z,-z,1-z,y)+12 G(0,0,0,1,y)-24 G(0,0,0,1,z)+8
   G(0,0,1,1,y)+8 G(0,0,1-z,1,y)-20 G(0,0,-z,1-z,y)+22 G(0,1,1,1,z)+12 G(0,1,1-z,1,y)+6 G(0,1-z,0,1,y)+8 G(0,1-z,1,1,y)-4
   G(0,1-z,1-z,1,y)+6 G(0,1-z,-z,1-z,y)+8 G(0,-z,0,1,y)+8 G(0,-z,1-z,1,y)+4 G(1-z,0,0,1,y)+8 G(1-z,0,1,1,y)-4
   G(1-z,0,1-z,1,y)-4 G(1-z,1-z,0,1,y)+6 G(1-z,-z,0,1,y)+6 G(1-z,-z,1-z,1,y)+16 G(-z,0,0,1,y)+8 G(-z,0,1-z,1,y)+8
   G(-z,1-z,0,1,y)+48 G(-z,1-z,1-z,1-z,y)+8 G(-z,-z,1-z,1-z,y)-40 G(-z,-z,-z,1-z,y)+\left(-G(1,z)^2-2 G(0,z) G(1,z)+4
   G(1-z,y)^2+G(1,y) (2 G(0,z)-2 G(1,z)-2 G(1-z,y))+4 G(0,z) G(1-z,y)+G(0,y) (-6 G(1,y)+6 G(1,z)+4 G(1-z,y))+6 G(0,1,y)-6
   G(0,1-z,y)-6 G(1-z,1,y)\right) \zeta (2)+\frac{93 \zeta (4)}{4}+(-22 G(1,y)-4 G(1,z)+18 G(1-z,y)) \zeta (3) \,,
\end{dmath}
\begin{dmath}
{\cal R}^{(2), C_F^2}_{\bar\psi \psi;4} = 
-\frac{4}{3} G(0,z) G(1,z)^3-2 G(0,z) G(1-z,y) G(1,z)^2+\frac{8}{3} G(-z,y)^3 G(1,z)+4 G(1,y) G(1-z,y) G(-z,y) G(1,z)+8
   G(0,0,-z,y) G(1,z)+\left(4 G(1,z)^2-4 G(0,z) G(1,z)\right) G(-z,y)^2-2 G(0,1,y)^2-2 G(0,1,z)^2+8 G(-z,1-z,y)^2+G(0,y)^2
   \left(2 G(1,y)^2+(-4 G(1,z)-4 G(1-z,y)) G(1,y)+2 G(1,z)^2+2 G(1-z,y)^2+4 G(1,z) G(1-z,y)\right)+G(1-z,y)^2 \left(2
   G(0,z)^2-8 G(1,z) G(-z,y)\right)+G(0,y) \left(-\frac{4}{3} G(1,y)^3-4 G(1,z) G(-z,y)^2-4 G(1,z)^2
   G(-z,y)\right)+\left(-8 G(1-z,y)^2+(4 G(0,z)-8 G(1,z)) G(1-z,y)+8 G(-z,y)^2+G(0,y) (4 G(1,y)-4 G(1,z)-4 G(1-z,y))+G(1,y)
   (4 G(1,z)+4 G(1-z,y))+8 G(1,z) G(-z,y)\right) G(0,1,z)+(-8 G(0,z) G(1,z)-4 G(-z,y) G(1,z)-4 G(0,z) G(1-z,y)+8 G(0,1,z))
   G(0,1-z,y)+(8 G(0,z) G(1,z)+4 G(1,y) G(1,z)-8 G(0,1,z)) G(0,-z,y)+G(1-z,1,y) (G(0,y) (4 G(1,y)+4 G(1,z)+4 G(1-z,y))-4
   G(1,z) G(-z,y)-4 G(0,1,z)-4 G(-z,1-z,y))+G(0,1,y) \left(4 G(1,y)^2+G(0,y) (-4 G(1,y)+4 G(1,z)+4 G(1-z,y))-4 G(1,z)
   G(-z,y)-4 G(0,1,z)+12 G(1-z,1,y)-4 G(-z,1-z,y)\right)+\left(-8 G(1-z,y)^2+16 G(1,z) G(1-z,y)-8 G(0,y) G(1,z)+G(1,y) (4
   G(1-z,y)-4 G(1,z))+16 G(1,z) G(-z,y)+8 G(0,1,z)\right) G(-z,1-z,y)+(4 G(1,z)-4 G(1,y)) G(0,0,1,z)-8 G(0,z)
   G(0,0,1-z,y)+(8 G(0,y)-8 G(1,y)) G(0,1,1,y)+(8 G(0,z)-8 G(1,y)+8 G(1,z)+20 G(1-z,y)-8 G(-z,y)) G(0,1,1,z)+(-4 G(1,y)-4
   G(1,z)-4 G(1-z,y)) G(0,1-z,1,y)+8 G(0,z) G(0,1-z,1-z,y)+(8 G(0,z)+4 G(1,y)+4 G(1,z)) G(0,-z,1-z,y)+(-4 G(1,y)-4 G(1,z)-4
   G(1-z,y)) G(1-z,0,1,y)-16 G(0,y) G(1-z,1,1,y)-8 G(0,y) G(1-z,1-z,1,y)+(8 G(0,y)+4 G(1,z)+4 G(1-z,y)) G(-z,0,1,y)+(8
   G(0,z)+4 G(1,z)) G(-z,0,1-z,y)+(8 G(0,y)+4 G(1,z)+4 G(1-z,y)) G(-z,1-z,1,y)+(-8 G(0,y)-8 G(0,z)-8 G(1,y)-16 G(1,z)+32
   G(1-z,y)) G(-z,1-z,1-z,y)+(-8 G(0,y)-8 G(0,z)-16 G(1,z)) G(-z,-z,1-z,y)-8 G(0,0,1,1,z)+8 G(0,0,-z,1-z,y)+8
   G(0,1,1,1,y)-16 G(0,1,1,1,z)-12 G(0,1,1-z,1,y)-8 G(0,1-z,1,1,y)+8 G(0,1-z,1-z,1,y)-4 G(0,1-z,-z,1-z,y)-8 G(0,-z,0,1,y)-8
   G(0,-z,1-z,1,y)-8 G(1-z,0,1,1,y)+8 G(1-z,0,1-z,1,y)+8 G(1-z,1-z,0,1,y)-4 G(1-z,-z,0,1,y)-4 G(1-z,-z,1-z,1,y)-16
   G(-z,0,0,1,y)-8 G(-z,0,1-z,1,y)-8 G(-z,1-z,0,1,y)-48 G(-z,1-z,1-z,1-z,y)-16 G(-z,-z,1-z,1-z,y)+16
   G(-z,-z,-z,1-z,y)+\left(4 G(1,y)^2-8 G(1-z,y)^2-8 G(1,z) G(1-z,y)-8 G(0,1,y)+8 G(0,1-z,y)+8 G(1-z,1,y)\right) \zeta
   (2)-22 \zeta (4)+(16 G(1,y)+8 G(1,z)-8 G(1-z,y)) \zeta (3) \,.
\end{dmath}
We can see that each term of ${\cal R}^{(2)}_{\bar\psi \psi;4}$ is quite different from the previous case of ${\cal R}^{(2), C_F^2}_{{\cal O}_0;4}$ (note that both results are expressed in the same set of independent functional basis). However, it is quite remarkable that the sum of the three terms is the same, as given in \eqref{eq:L2-psipsi-relation}.

Finally, we give the result of ${\cal R}^{(2)}_{{\cal O}_4;4}(1^q, 2^{\bar q}, 3^\pm)$ for the length-3 operator ${\cal O}_4\sim F_{\mu\nu}D^\mu(\bar\psi\gamma^\nu\psi)$. It also has the same color structure as previous case, and in the basis of 2d Harmonic polylogarithms we have: 
\begin{dmath}
{\cal R}^{(2), C_A^2}_{{\cal O}_4;4} = 
\zeta (2) \left(G(0,y) (3 G(1-z,y)-2 G(0,z)+3 G(1,z))-G(1-z,y)^2+(3 G(0,z)-4 G(1,z)) G(1-z,y)-2 G(0,1-z,y)-2
   G(1-z,1,y)-\frac{1}{2} G(0,y)^2-\frac{1}{2} G(0,z)^2-2 G(1,z)^2+3 G(0,z) G(1,z)-2 G(0,1,z)\right)+\zeta (3) (11 G(1-z,y)-6
   G(0,y)-6 G(0,z)+11 G(1,z))-3 G(1,z) G(0,0,-z,y)+\left(\frac{1}{4} G(0,z)^2+\frac{1}{2} G(1,z) G(0,z)\right)
   G(1-z,y)^2+\left(\frac{3}{2} G(0,z) G(1,z)^2-\frac{1}{2} G(0,z)^2 G(1,z)\right) G(1-z,y)+G(0,y)^2 \left(\frac{1}{4}
   G(1-z,y)^2+\left(\frac{1}{2} G(0,z)+\frac{1}{2} G(1,z)\right) G(1-z,y)-\frac{1}{4} G(0,z)^2+\frac{1}{2} G(1,z)
   G(0,z)+\frac{1}{4} G(1,z)^2\right)+G(0,y) \left(-G(0,z) G(1-z,y)^2+\left(\frac{1}{2} G(0,z)^2-2 G(0,z) G(1,z)\right)
   G(1-z,y)+\frac{1}{2} G(1,z) G(0,z)^2-G(1,z)^2 G(0,z)\right)+G(0,1,z) \left(G(0,y) (G(1-z,y)+G(1,z))-\frac{1}{2} G(1-z,y)^2-3
   G(1,z) G(1-z,y)-G(0,y)^2-\frac{3}{2} G(1,z)^2\right)+\left(G(0,z) G(1-z,y)-G(0,z)^2+2 G(1,z) G(0,z)-2 G(0,1,z)\right)
   G(0,1-z,y)+\left(-G(0,y)^2-2 G(0,1,y)\right) G(1-z,1,y)+G(0,0,1,z) (2 G(1-z,y)+G(0,y)+2 G(1,z))+G(0,z) G(0,0,1-z,y)+G(0,1,1,z)
   (3 G(1-z,y)-G(0,y)+3 G(1,z))-G(0,z) G(0,1-z,1-z,y)+2 G(0,y) G(1-z,1,1,y)+G(0,y) G(1-z,1-z,1,y)+2 G(0,0,1-z,1,y)-3
   G(0,0,-z,1-z,y)+2 G(0,1,1-z,1,y)+2 G(0,1-z,0,1,y)+2 G(0,1-z,1,1,y)-G(0,1-z,1-z,1,y)+2 G(1-z,0,0,1,y)+2
   G(1-z,0,1,1,y)-G(1-z,0,1-z,1,y)-G(1-z,1-z,0,1,y)+\frac{1}{2} G(0,z) G(1,z)^3-\frac{1}{4} G(0,z)^2 G(1,z)^2-3 G(0,0,0,1,z)-2
   G(0,0,1,1,z)-3 G(0,1,1,1,z)-\frac{27 \zeta (4)}{8} \,,
\end{dmath}
\begin{dmath}
{\cal R}^{(2), C_A C_F}_{{\cal O}_4;4} = 
\zeta (2) \left(G(0,y) (-2 G(1-z,y)-2 G(1,z))+(2 G(1,z)-2 G(0,z)) G(1-z,y)+2 G(0,1-z,y)+2 G(1-z,1,y)+G(0,y)^2+G(0,z)^2+G(1,z)^2-2
   G(0,z) G(1,z)+2 G(0,1,z)\right)+\zeta (3) (-38 G(1-z,y)+6 G(0,y)+6 G(0,z)-38 G(1,z))+G(0,y)^2 \left(-G(1,z)
   G(1-z,y)-\frac{1}{2} G(1-z,y)^2-\frac{1}{2} G(1,z)^2\right)-2 G(0,y) G(0,1,1,z)-2 G(0,y) G(1-z,1,1,y)+2 G(0,y)
   G(1-z,1-z,1,y)+\left(G(0,z) G(1,z)-\frac{1}{2} G(0,z)^2\right) G(1-z,y)^2+G(0,1,z) \left(G(0,y) (2 G(1-z,y)+2
   G(1,z))-G(1-z,y)^2+G(0,y)^2\right)+\left(2 G(0,z) G(1-z,y)+G(0,z)^2-2 G(1,z) G(0,z)+2 G(0,1,z)\right)
   G(0,1-z,y)+\left(G(0,y)^2+2 G(0,1,y)\right) G(1-z,1,y)+G(0,0,1,z) (-2 G(1-z,y)-4 G(0,y)-2 G(1,z))-4 G(0,z) G(0,0,1-z,y)+6
   G(1,z) G(0,0,-z,y)-2 G(0,z) G(0,1-z,1-z,y)-2 G(0,0,1-z,1,y)+6 G(0,0,-z,1-z,y)-2 G(0,1,1-z,1,y)-2 G(0,1-z,0,1,y)-2
   G(0,1-z,1,1,y)-2 G(0,1-z,1-z,1,y)-2 G(1-z,0,0,1,y)-2 G(1-z,0,1,1,y)-2 G(1-z,0,1-z,1,y)-2 G(1-z,1-z,0,1,y)+6 G(0,0,0,1,z)+2
   G(0,0,1,1,z)+\frac{73 \zeta (4)}{4} \,,
\end{dmath}
\begin{dmath}
{\cal R}^{(2), C_F^2}_{{\cal O}_4;4} = 
\zeta (3) (24 G(1-z,y)+24 G(1,z))-22 \zeta (4) \,.
\end{dmath}
We can see that the explicit functions are very different between the length-2 and length-3 cases. The fact that they both satisfy the same correspondence we provided in the main text strongly suggests the principle is universal.

\end{document}